\documentclass[debug]{rmaa}

\usepackage{paralist}

\usepackage{psfrag,color}

\usepackage[latin1]{inputenc}

\title{Thirteen years of Weather Statistics at \\ San Pedro Martir Observatory} 

\author{
  I. Plauchu-Frayn,\altaffilmark{1} 
  E. Colorado,\altaffilmark{1}
  M. Richer,\altaffilmark{1}
  and C. Herrera-Vázquez\altaffilmark{2}}

\altaffiltext{1}{Instituto de Astronomía, Universidad Nacional Autónoma de México (IA-UNAM).}
\altaffiltext{2}{Centro de Investigación Científica y Educación Superior de Ensenada (CICESE).}

\shortauthor{Plauchu-Frayn et~al.}
\shorttitle{Weather Statistics at OAN-SPM\@.}

\fulladdresses{
\item I. Plauchu-Frayn, E. Colorado and M. Richer: Instituto de Astronomía, Universidad Nacional Autónoma de México, 
Apartado Postal 106, 22800 Ensenada, B.C., México. (ilse, colorado, richer@astro.unam.mx)
\item C. Herrera-Vázquez: Centro de Investigación Científica y Educación Superior de Ensenada, 
Carr. Ensenada-Tijuana No. 3918, Zona Playitas, C. P. 22860, Ensenada, B.C., México (cherrera@cicese.edu.mx)
}

\listofauthors{I. Plauchu-Frayn,  E. Colorado,  M. Richer \& C. Herrera-Vázquez, C.}
\indexauthor{Plauchu-Frayn, I.}
\indexauthor{Colorado, E.}
\indexauthor{Richer, M.}
\indexauthor{Herrera-Vázquez, C.}

\resumen{Presentamos la estadística del clima para trece años de datos obtenidos con las estaciones 
meteorológicas en el Observatorio Astronómico Nacional en la Sierra San Pedro Mártir (OAN-SPM) durante 
el período 2007--2019. Estas estaciones meteorológicas incluyen sensores que miden temperatura, presión 
atmosférica, humedad relativa, precipitación y las condiciones del viento, entre otras variables 
climatológicas. Los valores medianos de la temperatura del aire son $10.3^{\circ}$\,C y $7.0^{\circ}$\,C, 
para el día y la noche, repectivamente. La humedad relativa sigue una variación estacional, siendo los 
meses de abril a junio los más secos, mientras que los meses de junio a septiembre son los más húmedos. 
Los valores medianos de la velocidad del viento son 11 y 14\,km hr$^{-1}$, 
para el día y la noche, respectivamente. Las direcciones del viento dominantes son el SSW y el Norte. 
Los vientos sostenidos son más fuertes por la noche y durante los meses de diciembre, enero y febrero.
Nuestros datos indican un promedio anual de precipitación de lluvia de 313\,mm, gran parte de la cual 
ocurre durante las tardes del verano en forma de tormentas eléctricas. }

\abstract{We present weather statistics for thirteen years of data gathered with the meteorological stations 
at Observatorio Astronómico Nacional in the Sierra San Pedro Mártir (OAN-SPM) over the period 2007--2019. 
These weather stations include sensors that measure temperature, atmospheric pressure, relative humidity, 
precipitation and wind conditions, among other climatological variables. 
The median values of the air temperature are $10.3^{\circ}$\,C and $7.0^{\circ}$\,C for daytime and nighttime, 
respectively.  The relative humidity follows a seasonal variation with April-June being the driest 
months while July-September being the most humid.  The median values for the sustained wind speed are 
11 and 14\,km hr$^{-1}$ for daytime and nighttime data, respectively. Preferred wind 
directions are SSW and North. Sustained winds are stronger at night and during December, January and February. 
Our data indicate an annual mean rain precipitation of 313\,mm, most of which occurs during the summer season 
as afternoon thunderstorms. }

\addkeyword{data analysis -- site testing}

\begin{document}
\maketitle

\section{Introduction}
\label{sec:intro}

The Observatorio Astronómico Nacional in the Sierra San Pedro Mártir (hereinafter OAN-SPM) is located on the top 
of Sierra San Pedro Mártir in Baja California, México (2800\,m, +31$^{\circ}$\,02''\, 40'\, N, 115$^{\circ}$\,28''\, 00'\, W). 
The site excels in sky darkness (\citealp{2016PASP..128c5004T}; \citealp{2017PASP..129c5003P})  and an absence 
of cloud cover, with approximately 70\% and 80\% photometric and spectroscopic time, respectively \citep{2007RMxAC..31...47T}. 
The median seeing measured at zenith at 5000\AA\ varies from 0.55\arcsec\ to 0.79\arcsec, depending upon
the study (\citealp{1998RMxAA..34...47E}; \citealp{2003RMxAC..19...41E}; \citealp{2008RMxAA..44..231B}; 
\citealp{2009PASP..121.1151S}; \citealp{2012MNRAS.426..635S}; \citealp{2019MNRAS.490.1397A}) with the most 
recent measurements favoring a median value near 0.79\arcsec.  Similar seeing seems to be the rule at nearby 
sites \citep{2008RMxAA..44..231B}. Atmospheric extinction is typically 0.13\,mag airmass$^{-1}$ in V band 
(\citealp{2001RMxAA..37..187S}). Due to these excellent atmospheric conditions and favorable location away 
from large urban areas, the OAN-SPM is an excellent site for optical and infrared facilities. 

In the early 1960's, a search was begun to find the best site for optical astronomical observations in Mexico 
\citep{1973Mercu...2....9M}.  The site was chosen, using, among other information, satellite photographs and on-site 
testing, as recounted by \citet{1972BOTT....6..215M}.  Subsequent studies during the first years of the observatory's 
operation validated the choice of the site (e.g., \citealp{1971PASP...83..401W}; \citealp{1972BOTT....6..215M};
\citealp{1977RMxAA...2...43A}). The OAN-SPM was formally inagurated in 1979.

\citet{1992RMxAA..24..179T} presented a study of ten years (1982-1992) of weather and observing statistics 
at OAN-SPM site, based upon data reported by the telescope operators. He finds that the mean relative humidity is 
54\,\% and that the best seasons in terms of cloudlessness and low humidity are spring and autumn. 
\citet{2003RMxAC..19...75T} and \citet{2007RMxAC..31...47T} updated these results, based upon data spanning 22 years 
(1982--2002), finding that the fraction of photometric nights is close to 70\,\% for the most recent data and the 
fraction of spectroscopic nights is approximately 80\,\%.

\citet{2003RMxAC..19...99M} present a four year study with a weather station installed at the 1.5\,m telescope. They 
estimated that air temperatures are in the range of $-15$ to $20 ^{\circ}$\,C, the relative humidity 
shows a seasonal dependence with short variations on short time scales, especially during summer nights. Also, this 
study reports a range of 733 to 753\,mb for the atmospheric pressure. Daytime and nighttime wind speeds have median 
values of 14 and 19\,km hr$^{-1}$ with strong winds coming from the SSW direction.  

\citet{2003RMxAC..19...90H} studied the precipitable water vapor above the site over an eight year period (1995--2002) 
by measuring the optical depth of the atmosphere at 210\,GHz, finding a clear peak corresponding to the summer 
``monsoon" as well as year-to-year variations, e.g., ``El Niño" events and the like.  Subsequent analysis  
(\citealp{2009RMxAA..45..161O}; \citealp{2010PASP..122..470O}) indicates a median precipitable water vapor below 
4\,mm, except in summer.  

Here, we continue the tradition of characterizing the site of the OAN-SPM by presenting an analysis of weather 
variables acquired with the observatory's weather stations.  Since 2006, the OAN-SPM installed weather stations 
and has operated them continuously as facility instruments. This study describes the data gathered 
in a consistent and continuous manner by the weather stations during the last thirteen years of operation of the 
observatory (2007-2019). These stations include sensors for measuring  air temperature, atmospheric pressure, 
relative humidity, wind conditions, water precipitation, solar radiation, evapotranspiration, and the UV index. 
The equipment used to measure weather conditions is described in \S~\ref{sec:data}, we present our results in 
\S~\ref{sec:results} and finally, our conclusions are given in \S~\ref{sec:conclu}.

\section{Weather data archive and analysis}
\label{sec:data}

The data were acquired with two sets of instrumentation.  From mid-2006 to mid-2013, a Davis Instruments weather 
station (hereinafter DI; \citealp{CI2007-04}) model Vantage Pro2 Plus was used.  From mid-2013, a Vaisala Weather 
Transmiter (hereinafter VWT) model WXT520 with a Vaisala WINDCAP sensor was installed at OAN-SPM\@. 
The DI weather station has a mechanical cup anemometer (instant readings: 2.5--3 secs.), while the 
Vaisala WINDCAP sensor of the VWT weather station is ultrasonic (instant readings: 0.25 secs.).
The weather data recorded by these stations is currently available from the OAN-SPM 
homepage\footnote{\texttt{http://www.astrossp.unam.mx}}. The basic data that we consider in this paper are air 
temperature, relative humidity, precipitation, atmospheric pressure and wind speed and direction.  All meteorological 
variables are monitored continuously. Every five minutes the mean value of the air temperature, 
relative humidity and atmosperic pressure and the total accumulated precipitation are recorded on the hard disk of 
the control computer. For the sustained wind speed, the mean value of all instant readings within a five minute 
interval is determined, while for the gust wind speed, the maximum value in this interval is determined. A systematic 
error of $\pm5^{\circ}$ degrees is estimated in the alignment process of the station with respect to the true 
North pole. In Table \ref{tab:stations} we present the parameter accuracies for each station as indicated by 
the manufacturer.

\begin{table}[!t]\centering
  \setlength{\tabnotewidth}{0.5\columnwidth}
  \tablecols{3}
  \setlength{\tabcolsep}{1\tabcolsep}
  \caption{Weather station parameter accuracy} \label{tab:stations}
 \begin{tabular}{lll}
    \toprule
    Parameter  &  DI                  &  VWT    \\
          \midrule
    Wind speed       & $\pm5\,\%$     ($<241$\,km hr$^{-1}$)   & $\pm3\,\%$  ($<136$\,km hr$^{-1}$)  \\
    Wind direction   & $\pm7^{\circ}$         & $\pm3^{\circ}$  \\
    Temperature      & $\pm0.5^{\circ}$\,C   ($<43^{\circ}$\,C)   & $\pm0.2-0.3^{\circ}$\,C   (-20 to 20$^{\circ}$\,C)  \\
    Humidity         & $\pm3\,\%$  (0 to 90\%)   & $\pm3\,\%$ (0 to 90\%)  \\
                     & $\pm4\,\%$  ($>90\%$)    & $\pm5\,\%$ ($>90\%$)   \\
    Pressure         & $\pm0.3$\,mb             & $\pm0.5$\,mb  (0 to 30$^{\circ}$\,C) \\
    Precipitation    & $\pm4\,\%$               & $\pm5\,\%$   \\

    \bottomrule
  \end{tabular}
\end{table}

\begin{table}[!t]\centering
  \begin{changemargin}{-2.5cm}{-2cm}
  \setlength{\tabnotewidth}{0.5\columnwidth}
  \tablecols{5}
  \setlength{\tabcolsep}{1\tabcolsep}
  \caption{Weather data archive} \label{tab:data}
 \begin{tabular}{lccll}
    \toprule
    Year & N of days & \% of the year & Location & Comments  \\
          \midrule
   2007   & $ 319 $   & $87.4\% $ &  site 1        & The barometer calibration was incorrect.\\
   2008   & $ 344 $   & $94.0\% $ &  site 1        & -- \\
   2009   & $ 365 $   & $100\%  $ &  site 1        & -- \\
   2010   & $ 351 $   & $96.2\% $ &  site 1        & -- \\
   2011   & $ 365 $   & $100\%  $ &  site 1        & The pluviometer was not working.\\
   2012   & $ 366 $   & $100\%  $ &  site 1        & Only 67\% of wind data.  The anemometer was broken. \\
   2013   & $ 350 $   & $95.9\% $ &  site 1 \& 2   & The wind direction is not reliable for site 2. \\
   2014   & $ 364 $   & $99.7\% $ &  site 2        & The wind direction is not reliable for site 2. \\
   2015   & $ 352 $   & $96.4\% $ &  site 2 \& 3   & The wind direction is not reliable for site 2. \\
   2016   & $ 363 $   & $99.2\% $ &  site 3        & -- \\
   2017   & $ 351 $   & $96.2\% $ &  site 3        & -- \\
   2018   & $ 363 $   & $99.4\% $ &  site 3        & -- \\
   2019   & $ 361 $   & $98.9\% $ &  site 3        & -- \\

    \bottomrule
  \end{tabular}
  \end{changemargin}
\end{table}

\begin{figure}[!ht]\centering
  \includegraphics[width=1.0\columnwidth]{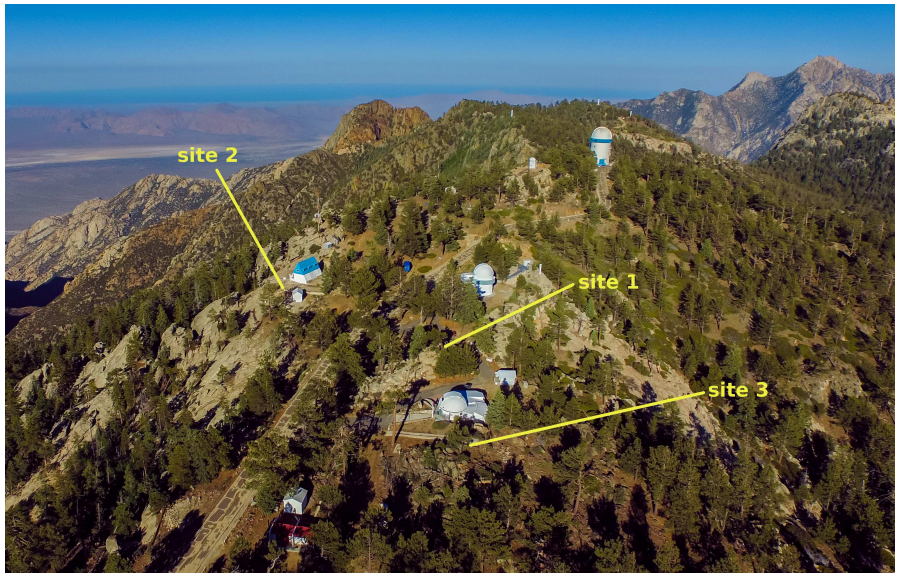}
  \caption{This aerial view of the OAN-SPM indicates the locations of the DI (site 1) and VWT (site 2 and later 
  site 3) weather stations during period 2006--2019.  The photo is courtesy of Eduardo López Ángeles.}
  \label{fig:sites}
\end{figure}

The two weather stations are quite similar, with the main difference being a better precision in the 
wind speed and direction for the VWT station. These two stations have been located at three different
sites within the OAN-SPM.  The DI station was located on a 6m mast erected upon a rock outcrop about 
3\,m high between the 1.5\,m and 0.84\,m telescopes (site 1). The VWT station was first installed on 
a 6\,m mast near the Cabaña Azul from August 2013 to November 2015 (site 2) and subsequently on a 6\,m 
mast to the west of the 1.5\,m telescope (site 3). These three different locations are shown in Figure~\ref{fig:sites}.

For this study, we retain thirteen whole years of data (2007-2019), representing $\sim$97\,\% of the time 
span available ($\sim$1.3 million 5-minute data points). Table~\ref{tab:data} presents a summary of the data. 
The principal gaps in the data are as follows.  There are gaps of approximately two weeks in December 2007 
and December 2008, corresponding to the staff holiday period, and there is a gap of several weeks in January 
2010, when the observatory was evacuated due to a severe snow storm.  During these periods, the weather station 
was switched off.  In addition, there were periods when some of the individual sensors were not functioning, 
as noted in Table \ref{tab:data}.  

Also, the data available concerning the precipitation during winter should be considered a lower limit.  The 
pluviometers for both weather stations are unreliable at sub-zero temperatures (they freeze) and under-record 
the precipitation that falls as snow.  We shall address this limitation in \S\ref{sec:prec}.

As a first step in our study, we compared each of the weather parameters collected by the weather stations at 
the three sites. The distributions of air temperature and relative humidity for both weather stations are 
similar with no shift between them. On the other hand, for the distribution of the atmospheric pressure, we find 
a shift of $\sim$1\,mb in the distribution of the VWT station compared to the DI  station. In addition, when 
we compare the distributions of the accumulated precipitation of two stations, we find similar distributions, 
with the VWT distribution slightly shifted to larger values compared to the DI distribution. Because these two 
weather stations never run simultaneously, we can not rule out that differences found are due to an increase 
of these variables over time. When we compare the distributions for the wind conditions, we found that the
only difference among sites was for the wind direction, which lead us to not use the wind direction data 
collected at site 2, a topic we shall address further in \S\ref{sec:wind}.

For the analysis of the data and presentation of the results, we have constructed two data sets: one for daytime 
and one for nighttime. The nighttime data set includes only data collected when the Sun was $\ge6 ^{\circ}$ below 
the horizon (i.e., the period between two consecutive Civil twilights) while the daytime data set consist of the 
data collected during the rest of the day. Both data sets, originally gathered in 5 minute intervals, have been 
reduced to hourly, daily, monthly, seasonal and annual means, medians, modes or sums depending upon the weather 
parameter or the issue to be studied. Finally, we define the seasons as follows: winter includes January, February, 
and March; spring, April, May, and June; summer, July, August, and September, and autumn, October, November, 
and December.

\section{Results}
\label{sec:results}

\subsection{Air temperature}
\label{sec:temp}

\begin{figure}[!ht]\centering
  \includegraphics[width=\columnwidth]{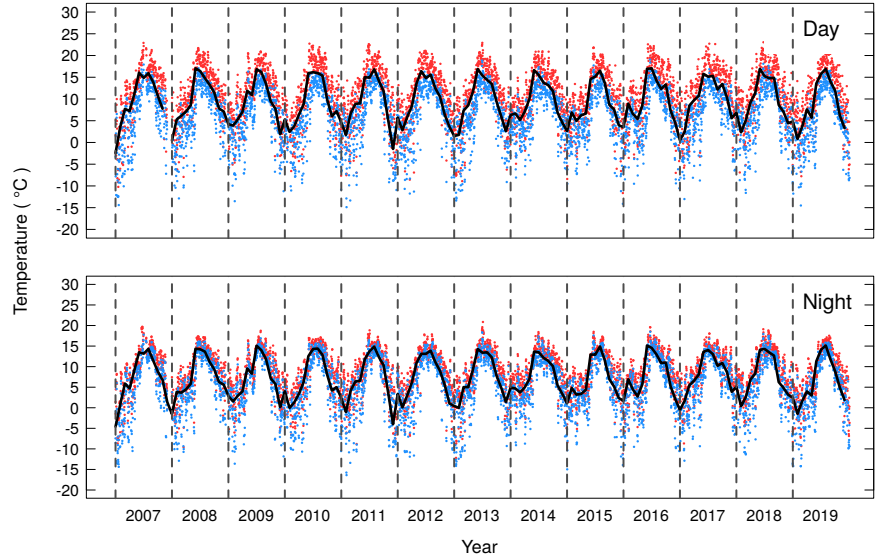}
  \caption{We present the daily maximum (red) and minimum (blue) air temperature for daytime and nighttime data 
  for the thirteen year time span of our data (2007--2019).  In each panel the black line represents the median 
  monthly air temperature, while the vertical dashed lines indicate the first day of each year.}
  \label{fig:temp}
\end{figure}

The daily maximum and minimum values of the air temperature from 2007 to 2019 are plotted in Figure~\ref{fig:temp}. 
The air temperature spans the range from $-17^{\circ}$\,C to $23^{\circ}$\,C over the course of the entire study 
period. From Figure~\ref{fig:temp}, it can be seen that the daytime temperatures cover a wider range than the 
temperatures at night. Also, there is a larger dispersion in the temperature during the winter (the valleys in 
the curves).

\begin{figure*}[!ht]\centering
  \newlength\thisfigwidth
  \setlength\thisfigwidth{0.5\linewidth}
  \addtolength\thisfigwidth{-0.5cm}
  \makebox[\thisfigwidth][l]{\textbf{a}}%
  \hfill%
  \makebox[\thisfigwidth][l]{\textbf{b}}\\[-3ex]
  \parbox[t]{\linewidth}{%
     \vspace{0pt}
     \includegraphics[width=6cm,height=6cm]{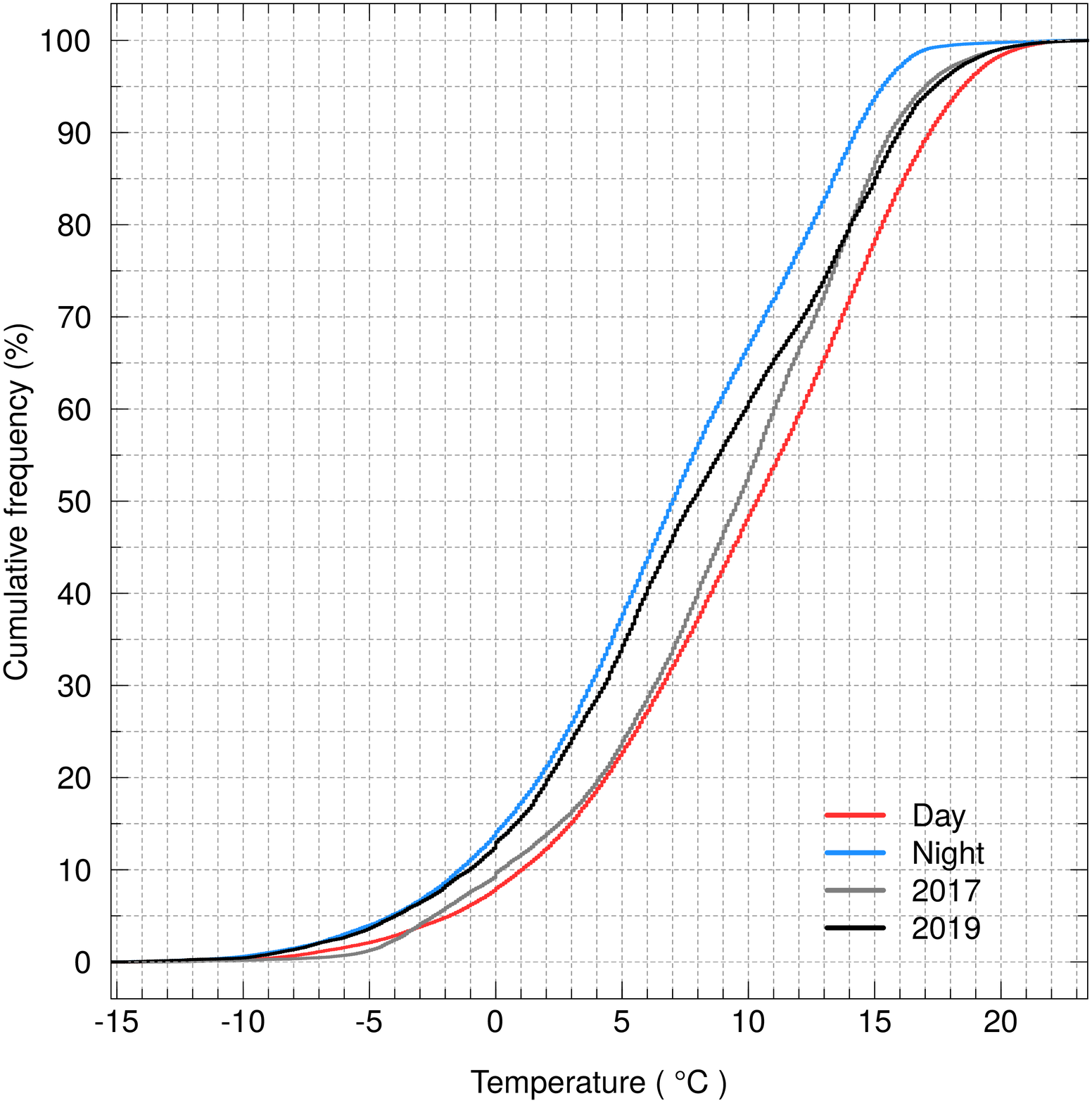}%
     \hfill%
     \includegraphics[width=6cm,height=6cm]{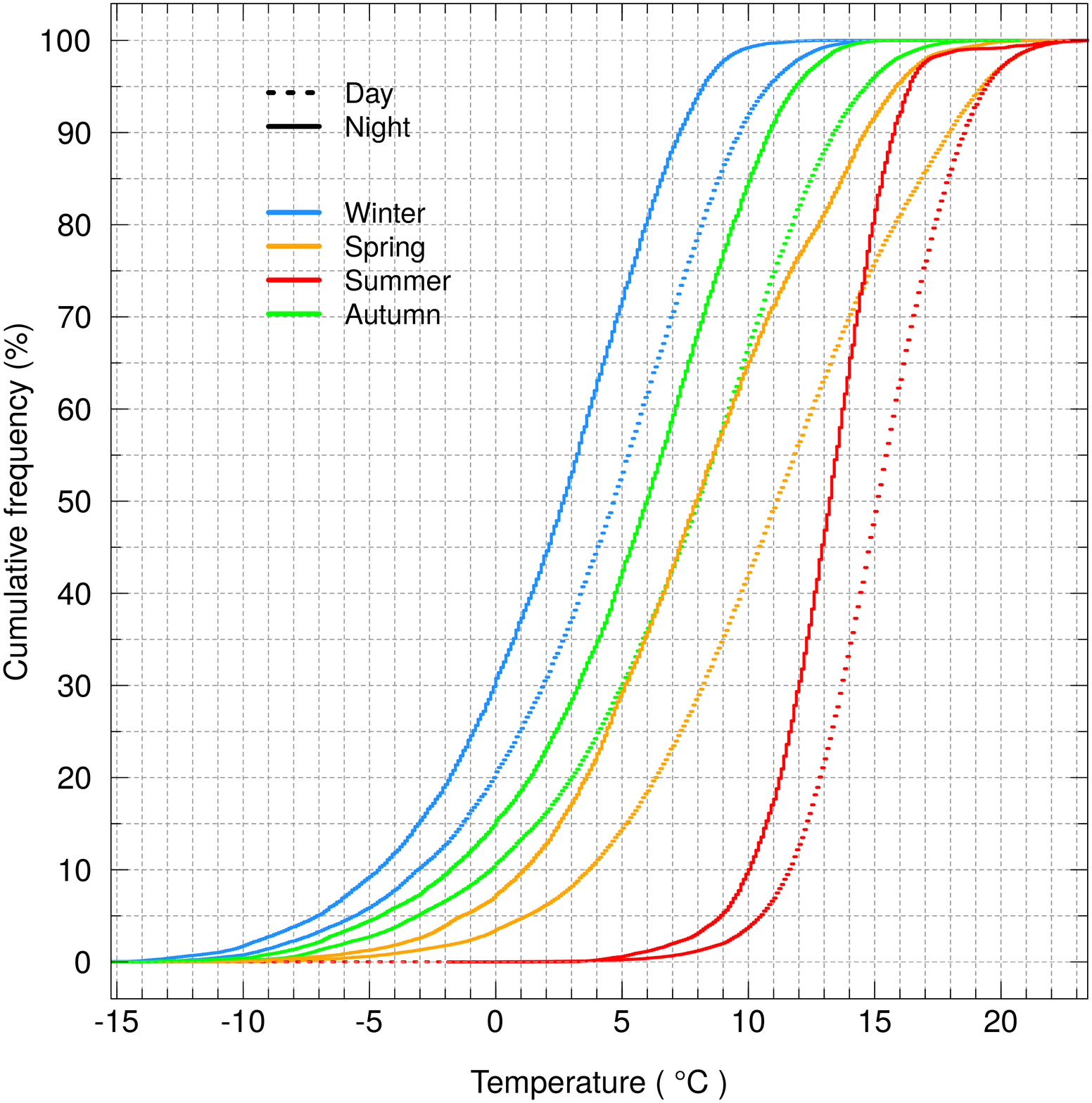}
     }
  \caption{We present (\textit{a})  the cumulative distribution of the daytime and nighttime air temperature 
  obtained in the period 2007--2019, based upon the 5 minute averages.  The blue line is for nighttime data, 
  red line is for daytime data, gray line is for 2017 (warmest year, day+night) data and black line is for 2019 
  data (coldest, day+night). In (\textit{b}) we present the cumulative distribution of air temperature by season 
  in the same period for daytime (dotted line) and nighttime (solid line). The blue line is for winter, 
  orange line for spring, red line for summer and green line for autumn. In both plots, 
  the x-axis does not cover the extremes of total temperature range (-17 to 23$^{\circ}$\,C). Values 
  below -10$^{\circ}$\,C and above 20$^{\circ}$\,C represent only 0.4\% and 1.0\% of the data, respectively.}
  \label{fig:tempD}
\end{figure*}

The cumulative distribution of daytime and nighttime air temperatures based upon 5 minute averages is shown in 
Figure~\ref{fig:tempD}a. The median daytime and nighttime air temperatures are $10.3^{\circ}$\,C and $7.0^{\circ}$\,C, 
respectively. From Figure~\ref{fig:tempD}a, it can be seen that 90\,\% of the time the air temperature is below 
$17^{\circ}$\,C, both day and night. The minimum temperature registered is $-16.7^{\circ}$\,C, with 
temperatures below $0^{\circ}$\,C and $-4^{\circ}$\,C occuring only 11\% and 4\% of the time, respectively. 
In Figure~\ref{fig:tempD}b, we present the cumulative distribution of air temperatures by season for daytime and 
nighttime. From Figure~\ref{fig:tempD}b, it can be seen that the warmest season is summer with a median value of the air 
temperature of $15.1^{\circ}$\,C and $13.2^{\circ}$\,C, for daytime and nighttime, respectively. The coldest season is 
winter with median values of $4.7^{\circ}$\,C  and $2.7^{\circ}$\,C, for daytime and nighttime, respectively. 
Spring is the season with the largest difference between daytime and nighttime temperatures, with median values of 
the air temperature $11.1^{\circ}$\,C and $7.9^{\circ}$\,C, respectively. In autumn, median values of the air 
temperature are $8.0^{\circ}$\,C for daytime and $6.0^{\circ}$\,C, for nighttime.

\begin{figure}\centering
  \includegraphics[width=1.0\columnwidth]{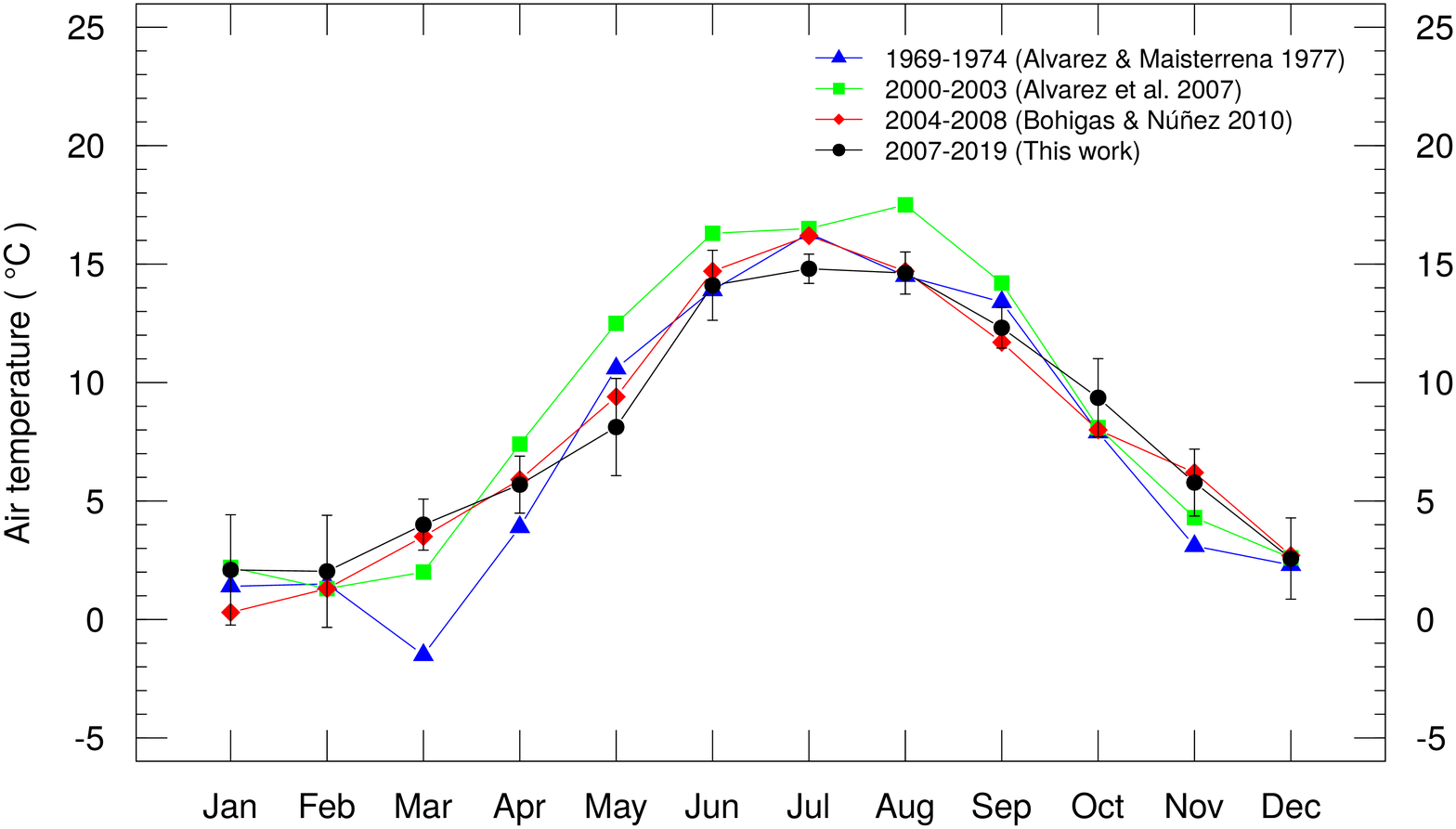}
  \caption{We compare the monthly mean of the air temperatures from this study with data from 
  \citet{1977RMxAA...2...43A}, \citet{2007RMxAC..31..113A} and \citet{2010RMxAA..46...89B} and this work.  
  The agreement is generally very good.}
  \label{fig:TempComp}
\end{figure}

We present the monthly means of the air temperature for daytime and nighttime in Table~\ref{tab:temp} and later in Figure~\ref{fig:allparam} and for each year, the monthly medians are presented in Table~\ref{tab:tempMY}. The mean 
temperature at the OAN-SPM has a very clear seasonal variation as can also be seen in Figure~\ref{fig:tempD}b.  
The warmest months are June, July and August and the coldest are December, January and February. The variation between 
the warmest and coldest months is $13^{\circ}$\,C (see Table~\ref{tab:temp}). The dispersion about the mean temperatures 
in Table~\ref{tab:temp} confirms the impression from Figure \ref{fig:temp} that there is more dispersion in temperature 
in winter and during the day.  In Figure~\ref{fig:TempComp}, we compare the monthly mean of the air temperatures obtained 
here with those from \citet{1977RMxAA...2...43A}, \citet{2007RMxAC..31..113A}, and \citet{2010RMxAA..46...89B}. All of the 
data sets are quite similar, except for one point in the 1969--1974 data set, due to a very cold March in 1973  ($-3.1^{\circ}$\,C; see Table 1 from \citealp{2007RMxAC..31..113A}).

\begin{table}[!ht]\centering
  \setlength{\tabnotewidth}{0.5\columnwidth}
  \tablecols{4}
  \setlength{\tabcolsep}{1\tabcolsep}
  \caption{Monthly mean air temperature} \label{tab:temp}
 \begin{tabular}{lrrr}
    \toprule
     & Day       & Night         &  $\mathrm{Day}+\mathrm{Night}$ \\
   \cmidrule{2-4}
   Month  & \multicolumn{3}{c}{($^{\circ}$\,C)}\\
   \cmidrule{1-4}
    January    & $ 2.9\pm  4.7$   & $  1.3\pm  4.4$   & $ 2.1\pm  4.5$\\
    February   & $ 3.0\pm  5.0$   & $  1.1\pm  4.8$   & $ 2.1\pm  4.9$\\
    March      & $ 5.1\pm  4.1$   & $  3.0\pm  3.9$   & $ 4.2\pm  4.0$\\
    April      & $ 6.9\pm  4.1$   & $  4.4\pm  3.8$   & $ 6.0\pm  3.9$\\
    May        & $ 9.4\pm  4.2$   & $  6.8\pm  4.1$   & $ 8.5\pm  4.1$\\
    June       & $ 15.4\pm 2.9$   & $  12.8\pm 3.0$   & $ 14.5\pm 2.9$\\
    July       & $ 15.7\pm 1.8$   & $  13.9\pm 1.7$   & $ 15.1\pm 1.7$\\
    August     & $ 15.5\pm 1.8$   & $  13.8\pm 1.6$   & $ 14.8\pm 1.7$\\
    September  & $ 13.3\pm 2.3$   & $  11.3\pm 2.1$   & $ 12.5\pm 2.2$\\
    October    & $ 10.4\pm 3.3$   & $  8.3\pm  3.2$   & $ 9.4\pm  3.2$\\
    November   & $ 6.5\pm  4.2$   & $  5.0\pm  3.8$   & $ 5.9\pm  4.0$\\
    December   & $ 3.3\pm  5.1$   & $  1.8\pm  4.7$   & $ 2.5\pm  4.9$\\

    \bottomrule
  \end{tabular}
\end{table}

\begin{table}[!ht]\centering
  \begin{changemargin}{-2cm}{-2cm}
  \setlength{\tabnotewidth}{1.0\columnwidth}
  \tablecols{14}
  \setlength{\tabcolsep}{1.1\tabcolsep}
  \caption{Monthly medians of air temperature} \label{tab:tempMY}
 \begin{tabular}{lrrrrrrrrrrrrr}
    \toprule
           & 2007 & 2008 & 2009 & 2010 & 2011 & 2012 & 2013 & 2014 & 2015 & 2016 & 2017 & 2018 & 2019 \\
   \cmidrule{2-14}
   Month   & \multicolumn{12}{c}{($^{\circ}$\,C)}  \\
   \cmidrule{1-14}
    Jan    & $-3.4$                 & $-0.9$\tabnotemark{b} & $3.6 $    & $ 4.2$\tabnotemark{d} & $3.4 $ & $4.2 $ & $0.9 $ & $5.5 $ & $1.4 $                 & $2.5 $ & $ 0.1$ & $6.1 $  & $3.3 $  \\
    Feb    & $1.9 $                 & $ 4.8$                & $2.6 $    & $ 1.1$                & $0.7 $ & $1.8 $ & $1.0 $ & $5.9 $ & $6.0 $                 & $8.1 $ & $ 1.9$ & $1.4 $  & $-0.1 $  \\
    Mar    & $6.7 $                 & $ 5.2$                & $4.2 $    & $ 2.7$                & $5.9 $ & $4.7 $ & $6.2 $ & $4.3 $ & $4.3 $                 & $5.6 $ & $ 7.1$ & $3.4 $  & $2.6 $  \\
    Apr    & $6.1 $                 & $ 6.0$                & $5.9 $    & $ 5.3$                & $7.8 $ & $7.6 $ & $6.9 $ & $6.3 $ & $5.2 $                 & $4.4 $ & $ 8.7$ & $8.4 $  & $6.1 $  \\
    May    & $10.3$                 & $ 7.5$                & $11.0$    & $ 7.9$                & $8.3 $ & $12.7$ & $11.3$ & $8.9 $ & $5.8 $                 & $7.0 $ & $ 9.9$ & $10.4$  & $4.5 $  \\
    Jun    & $15.2$                 & $16.0$                & $10.0$    & $14.7$                & $14.0$ & $15.3$ & $16.0$ & $15.7$ & $14.0$                 & $16.4$ & $15.0$ & $16.0$  & $12.9$  \\
    Jul    & $14.3$\tabnotemark{a}  & $15.7$                & $16.0$    & $15.4$                & $14.5$ & $14.2$ & $15.2$ & $14.8$ & $14.5$                 & $16.4$ & $14.7$ & $14.8$  & $15.2$  \\
    Aug    & $15.4$                 & $14.5$                & $15.3$    & $15.3$                & $16.1$ & $14.9$ & $14.3$ & $13.2$ & $16.1$\tabnotemark{e}  & $13.7$ & $14.2$ & $14.4$  & $16.3$  \\
    Sep    & $13.0$                 & $12.3$                & $12.8$    & $14.4$                & $13.1$ & $11.8$ & $13.0$ & $12.1$ & $12.9$                 & $11.4$ & $11.1$ & $14.1$  & $12.9$  \\
    Oct    & $10.1$                 & $10.7$                & $8.3 $    & $9.1 $                & $10.9$ & $10.0$ & $8.6 $ & $10.8$ & $8.2 $                 & $12.6$ & $12.2$ & $7.6 $  & $10.5$  \\
    Nov    & $7.3 $                 & $7.1 $                & $6.9 $    & $5.1 $                & $4.1 $ & $5.9 $ & $5.6 $ & $6.0 $ & $6.2 $                 & $6.5 $ & $9.6 $ & $6.0 $  & $5.3 $  \\
    Dec    & \nodata                & $6.2 $\tabnotemark{c} & $0.8 $    & $6.3 $                & $-3.0$ & $2.2 $ & $1.8 $ & $3.5 $ & $2.9 $                 & $3.1 $ & $4.8 $ & $3.5 $  & $2.4 $  \\
\midrule
  Total    & $9.6 $\tabnotemark{f}  & $8.7 $\tabnotemark{f} & $8.2 $    & $8.2 $                & $8.4 $ & $9.0 $ & $8.4 $ & $8.8 $ & $7.6 $                 & $8.4 $ & $9.4 $ & $8.4 $  & $7.5 $  \\
    Day    & $10.6$                 & $9.6 $                & $9.3 $    & $9.3 $                & $9.7 $ & $9.9 $ & $9.3 $ & $9.9 $ & $8.5 $                 & $9.5 $ & $10.7$ & $9.3 $  & $8.4 $  \\
  Night    & $8.2 $                 & $7.5 $                & $6.7 $    & $7.1 $                & $7.1 $ & $7.8 $ & $7.4 $ & $8.0 $ & $6.6 $                 & $7.2 $ & $8.2 $ & $7.3 $  & $6.0 $  \\
    \bottomrule
    \tabnotetext{a}{\small No data for the 3--17th of the month.}
    \tabnotetext{b}{\small No data for 10--15th of the month.}
    \tabnotetext{c}{\small No data for the 14--28th of the month.}
    \tabnotetext{d}{\small No data for the 20--30th of the month.}
    \tabnotetext{e}{\small No data for 26--31th of the month.}
    \tabnotetext{f}{\small $< 95\%$ of data (see  Table~\ref{tab:data}).}

  \end{tabular}
 \end{changemargin}
\end{table}

In the last three rows of Table~\ref{tab:tempMY} we report the median annual total ($\mathrm{day}+\mathrm{night}$), 
daytime and nighttime  air temperature, which are determined as the median value of the distribution of 
the daily mean temperatures for each year. By this measure, 2017 was the warmest year in the last thirteen 
years (see Figure~\ref{fig:tempD}a) with daytime and nighttime median air temperatures of $10.7^{\circ}$\,C 
and $8.2^{\circ}$\,C, respectively. The coldest year was 2019 (see Figure~\ref{fig:tempD}a) with daytime and 
nighttime median air temperatures of $8.4^{\circ}$\,C and $6.0^{\circ}$\,C respectively. Note that 2007 and 2008 
have missing data for coldest months, so their median temperatures are likely biased to slightly warmer 
temperatures than the true values.

\begin{figure}[!ht]\centering
  \includegraphics[width=0.7\columnwidth]{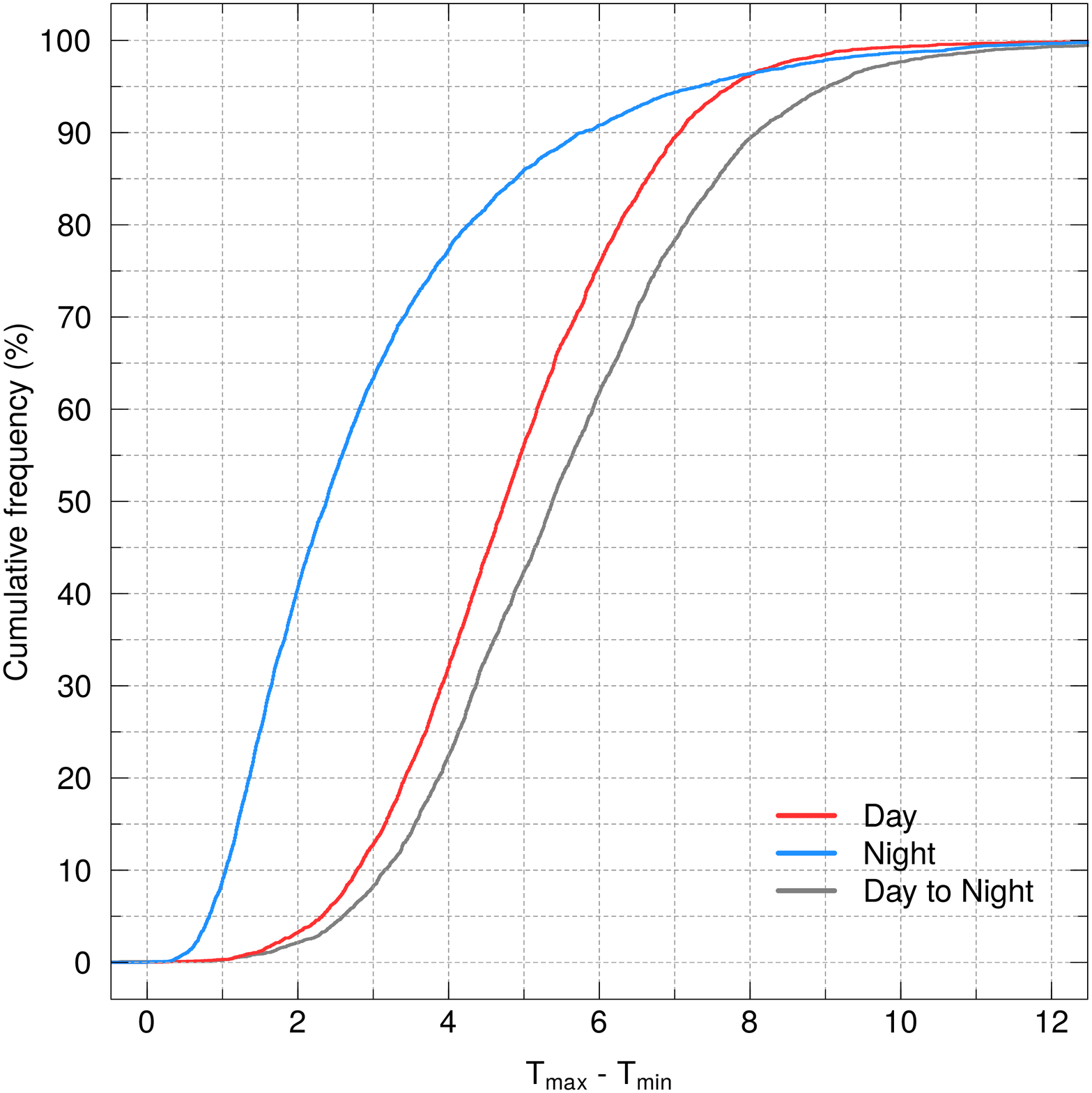}
    \caption{We present the cumulative distributions of difference between the maximum and minimum air 
    temperatures during an entire day (gray), during daytime (red), and nighttime (blue).  Clearly, the 
    temperature varies substantially less at night than during the day.}  \label{fig:tempdiff}
\end{figure}

In Figure~\ref{fig:tempdiff}, we show the cumulative distribution of the daily daytime, 
nighttime, and day-to-night air temperature variations. We define these temperature variations as the 
difference between the maximum and minimum temperatures during daytime, nighttime and over 
24 hour day. In a single day, maximum air temperature variation is smaller than $8^{\circ}$\,C 
90\,\% of the time with a daily median air temperature variation of $5.4^{\circ}$\,C. During daytime 
this variation is larger than at night, with 
median values of $4.8^{\circ}$\,C and $2.4^{\circ}$C for the day and night, respectively.

\begin{figure}[!ht]\centering
  \includegraphics[width=1.0\columnwidth]{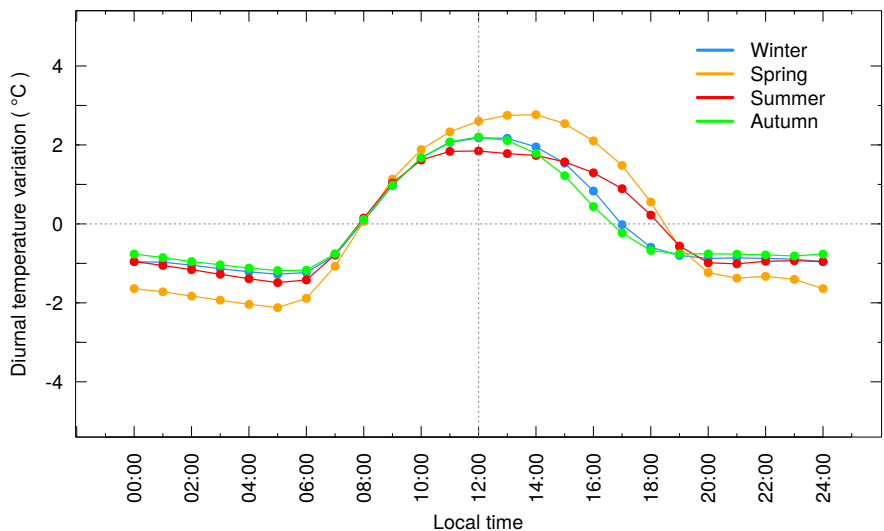}
  \caption{We present the diurnal air temperature variation by season.  To construct these curves, the mean 
  temperature   for each entire day (24 hours) is subtracted from the mean temperature for each hour and the 
  values for the days in a   given season are averaged. Again, we see that there is little temperature variation 
  at night.  The greatest diurnal temperature  variation occurs in spring.}
  \label{fig:tempH}
\end{figure}

Figure~\ref{fig:tempH} presents the diurnal variation of the temperature by season. For each season, we 
determine the mean temperature of the entire day (24 hours) and subtract this temperature from the temperature 
at a given hour. Values below zero indicate that the temperature is cooler than the daily mean temperature. 
During daytime ($\sim$06:00 to 12:00 hrs), the air temperature variation is large, typically 
$\Delta T \sim 3^{\circ}$\,C, with the most extreme variation in the spring, ($\Delta T \sim 5^{\circ}$\,C). 
The air is heated after sunrise with the most rapid phase occurring before 10:00 local time.  There is less 
variation during the middle 4-6 hours of the day. The last four hours before sunset see rapid cooling. 
From Figure~\ref{fig:tempH}, we can se that during the night the air temperature varies little in all seasons. 

\begin{figure}[!ht]\centering
  \includegraphics[width=1.0\columnwidth]{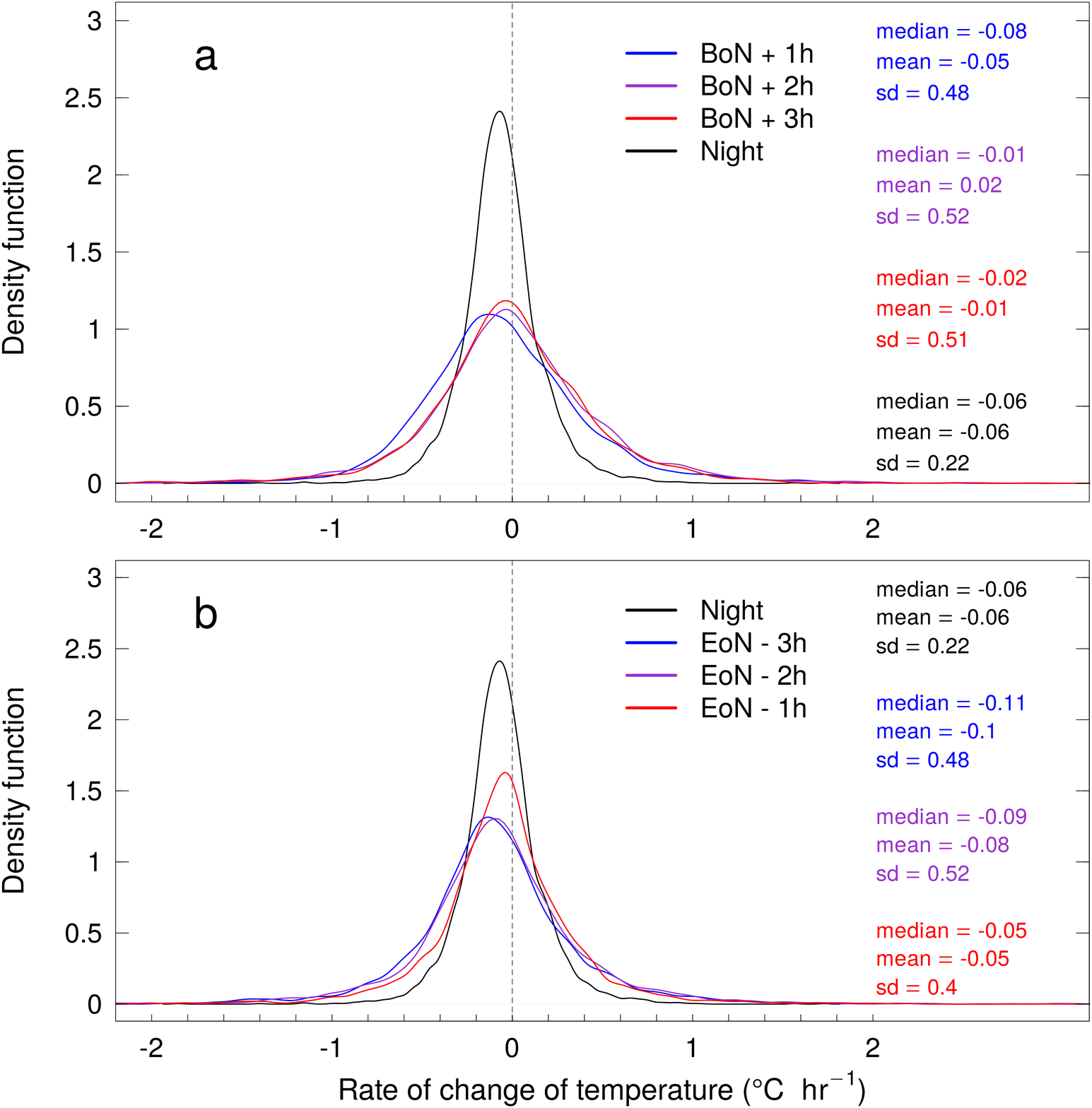}
  \caption{ We present the kernel density plots of the rate of change of the air temperature for: a) 
  1h, 2h, 3h after the beginning of the night (BoN) and the period of time including 3 hours after BoN and 3 hours 
  before the end of the night (EoN) or simply ''Night`` and b) 1h, 2h, 3h before the end of night and the 
  ''Night`` period. En each panel, we indicate the median, mean, and standard deviation for each distribution 
  of rates of change with the same color as its respective curve. Negative values of the rates of change of 
  the air temperature gradient indicate cooling, while positive rates indicate heating.}
  \label{fig:tempVar}
\end{figure}

In order to estimate how little air temperature varies during the night, we have estimated the rate of 
change of the air temperature during nightime. These rates of change are defined as the air temperature at a given 
hour minus the air temperature an hour before. According to this definition, positive rates indicate heating, 
while negative rates indicate cooling. In Figure~\ref{fig:tempVar}a, we present the density functions of the rate 
of change of the air temperature for 1, 2, 3 hours after the beginning of the night (BoN; Sun is $6^{\circ}$ below 
the horizon, i.e., 24 minutes after sunset) and the period of time between three hours after BoN and three hours 
before the end of the night (EoN, i.e., 24 minutes before sunrise). We refer to this period simply as ''Night`` in 
Figure~\ref{fig:tempVar}a and b. From Figure~\ref{fig:tempVar}a, it can be seen that after the first hour after the 
BoN there is a rapid cooling with median value of $-0.08^{\circ}$\,C hr$^{-1}$. Later, the cooling continues in the 
second and third hour after the BoN ($\sim-0.015^{\circ}$\,C hr$^{-1}$), although it is not as rapid as in the 
first hour after de BoN. In the ''Night`` period, the cooling has a median value of $-0.06^{\circ}$\,C hr$^{-1}$. 
In addtion, in Figure~\ref{fig:tempVar}b, we present the density functions of the air temperature gradients 1, 2, 3 
hours before the end of the night (EoN) and the ''Night`` period previously defined. From Figure~\ref{fig:tempVar}b, 
it can be seen that in three hours before the EoN, the cooling is the most rapid with a median value of 
$-0.11^{\circ}$\,C hr$^{-1}$. This cooling continues with a median value of $-0.09^{\circ}$\,C hr$^{-1}$ two 
hours before the EoN and $-0.05^{\circ}$\,C hr$^{-1}$ one hour before the EoN. Overall, the air temperature gradually 
decreases throughout the night at a mean rate of $-0.06^{\circ}$\,C hr$^{-1}$, with the most rapid cooling one hour 
after the BoN and the three hours before the EoN.

In Table~\ref{tab:tempCold}, we present the dates with the lowest temperatures recorded in the last thirteen 
years.  We have searched for dates where the daytime or nighttime mean temperature was $\le-10^{\circ}$\,C. From 
this table it can be seen the lowest temperatures occur from December to March, usually at night and when the 
relative humidity is high, perhaps due to the presence of sleet and/or snow. Such conditions are unusual, since 
the median relative humidity in Table~\ref{tab:tempCold} is 70\%, which occurs less than 15\% of the time (see \S \ref{sec:hum}).  

\begin{table}[!ht]\centering
  \setlength{\tabnotewidth}{0.5\columnwidth}
  \tablecols{5}
  \setlength{\tabcolsep}{1.5\tabcolsep}
  \caption{Lowest mean temperatures} \label{tab:tempCold}
  \begin{tabular}{rccccc}
    \toprule
         &   Day/Night   & \multicolumn{2}{c}{Temperature}   & RH    \\
         &                               &  mean   &  min            &  mean \\
   \cmidrule{2-5}
   \multicolumn{1}{c}{Date} &                          & \multicolumn{2}{c}{($^{\circ}$\,C)} & (\%)       \\
   \cmidrule{1-5}
   Jan 13, 2007      & D     & $-10.2$  & $-12.2$  & $80 $   \\
   Jan 13, 2007      & N     & $-10.4$  & $-13.6$  & $31 $   \\
   Jan 15, 2007      & D     & $-10.2$  & $-12.5$  & $36 $   \\
   Jan 18, 2007      & N     & $-10.2$  & $-12.7$  & $69 $   \\
   Jan 19, 2007      & D     & $-11.1$  & $-12.4$  & $92 $   \\
   Jan 19, 2007      & N     & $-11.6$  & $-13.2$  & $97 $   \\
   Jan 22, 2007      & D     & $-10.7$  & $-14.4$  & $82 $   \\
   Jan 22, 2007      & N     & $-11.1$  & $-14.4$  & $78 $   \\
   Mar 16, 2008      & N     & $-10.0$  & $-10.7$  & $89 $   \\
   Feb  2, 2011      & N     & $-10.5$  & $-15.8$  & $21 $   \\
   Feb  3, 2011      & D     & $-10.3$  & $-14.9$  & $6  $   \\
   Feb  3, 2011      & N     & $-10.9$  & $-16.4$  & $13 $   \\
   Mar 18, 2012      & N     & $-10.7$  & $-11.9$  & $86 $   \\
   Jan 11, 2013      & N     & $-13.1$  & $-14.1$  & $75 $   \\
   Jan 12, 2013      & D     & $-10.3$  & $-13.4$  & $70 $   \\
   Jan 12, 2013      & N     & $-11.3$  & $-13.7$  & $66 $   \\
   Jan 14, 2013      & N     & $-11.0$  & $-13.1$  & $26 $   \\
   Dec 31, 2014      & D     & $-10.8$  & $-13.3$  & $84 $   \\
   Dec 26, 2015      & D     & $-12.8$  & $-14.1$  & $50 $   \\
   Dec 26, 2015      & N     & $-12.9$  & $-13.9$  & $47 $   \\
   Dec 29, 2018      & N     & $-10.5$  & $-13.9$  & $37 $   \\
   Jan  1, 2019      & N     & $-10.7$  & $-14.2$  & $64 $   \\
   Feb 18, 2019      & N     & $-10.1$  & $-14.2$  & $84 $   \\
   Feb 22, 2019      & N     & $-11.3$  & $-12.3$  & $83 $   \\
    \bottomrule

   \end{tabular}
\end{table}

Table~\ref{tab:tempHot} presents the dates where the daytime mean temperature was $\ge20^{\circ}$C. As 
expected, all of these extreme temperatures take place in June and July during daytime.  When these 
temperatures occur, the relative humidity is typically below its median value (see \S \ref{sec:hum}).

\begin{table}[!ht]\centering
  \setlength{\tabnotewidth}{0.5\columnwidth}
  \tablecols{5}
  \setlength{\tabcolsep}{1.5\tabcolsep}
  \caption{Highest mean temperatures} \label{tab:tempHot}
  \begin{tabular}{rcccc}
    \toprule
           & Day/Night  & \multicolumn{2}{c}{Temperature}   & RH     \\
           &                               &  mean   &  max                    &  mean \\
   \cmidrule{2-5}
   \multicolumn{1}{c}{Date}   &                        & \multicolumn{2}{c}{($^{\circ}$\,C)} & (\%)   \\
   \cmidrule{1-5}
   Jul 1, 2007     & D   & $20.4 $  & $22.9 $  & $18 $ \\
   Jul 2, 2007     & D   & $20.5 $  & $22.2 $  & $17 $ \\
   Jul 2, 2011     & D   & $20.0 $  & $22.4 $  & $19 $ \\
   Jun 28, 2013    & D   & $20.2 $  & $22.2 $  & $31 $ \\
   Jun 29, 2013    & D   & $21.2 $  & $22.5 $  & $24 $ \\
   Jun 30, 2013    & D   & $20.8 $  & $22.9 $  & $27 $ \\
   Jun 20, 2015    & D   & $20.1 $  & $21.8 $  & $10 $ \\
   Jun 19, 2016    & D   & $20.9 $  & $22.1 $  & $17 $ \\
   Jun 20, 2016    & D   & $20.5 $  & $22.1 $  & $24 $ \\
   Jul 7,  2017    & D   & $20.0 $  & $22.7 $  & $35 $ \\
   Jun 22, 2018    & D   & $20.5 $  & $23.1 $  & $9 $ \\
    \bottomrule    
 \end{tabular}
\end{table}

Studies of global warming (\citealp{IPCC2018}\footnote{The Intergovernmental Panel on Climate Change. 
\texttt{https://www.ipcc.ch/sr15/}}), performed over long periods of time ($\sim$30 years to correct for 
short-term natural fluctuations), have shown an increase in the temperature of 0.1--0.3$^{\circ}$\,C per 
decade. However, this value has been found to be higher in Earth's land regions than at sea, because oceans 
lose more heat by evaporation. Also, higher values are measured in the Northern Hemisphere compared to the 
Southern Hemisphere, because the former has more land surface to absorb more sunlight and more heat. Similarly, 
there is a seasonal dependence, with higher values for colder seasons. Some Northern Hemisphere mid-latitud 
winter locations are experiencing regional warming more than double the global average. 

\citet{2007RMxAC..31..113A} found an increase in temperature at the OAN-SPM site. They found that temperature 
increased by 1.3$^{\circ}$\,C from 1969--1974 to 2000-2003 (or 0.4$^{\circ}$\,C per decade).  We searched 
for a trend in temperature as a function of time in our data. The best linear fit to mean yearly temperature 
data in the 2009-2019 period does not give a significant variation of the temperature 
at a 95\% confidence level.  We present the coefficients of the linear regression fit in Table~\ref{tab:fitsVar}. 
(We do not include the years 2007 and 2008 since their temperatures are likely biased, as noted earlier.) If 
we consider data from  \citet{2007RMxAC..31..113A} and perform a linear regression fit to the data for 
1969--1974, 2000--2003, and 2009--2019 (a 50 year interval), we obtain an increase in temperature of 
$\sim 1^{\circ}$C (or $0.2^{\circ}$\,C per decade see Table~\ref{tab:fitsVar}), but this variation is also 
not significant at the 95\% confidence level. 

\begin{table}[!t]\centering
  \begin{changemargin}{-2cm}{-2cm}
  \setlength{\tabnotewidth}{0.5\columnwidth}
  \tablecols{4}
  \setlength{\tabcolsep}{1.5\tabcolsep}
    \caption{Parameter variations (2009--2019).} \label{tab:fitsVar}
 \begin{tabular}{lrrlr}
    \toprule
                        & \multicolumn{2}{c}{Coeff.\tabnotemark{a}}   &  \multicolumn{1}{c}{P}    & \multicolumn{1}{c}{Variation}   \\
   \cmidrule{2-5}
     Parameter          & \multicolumn{1}{c}{a\,$\pm\, \sigma$} & \multicolumn{1}{c}{b\,$\pm\, \sigma$}   &         &  (per decade) \\
   \cmidrule{1-5}
 Temperature                  & $7.94\pm 0.30$ ($^{\circ}$\,C)     & $0.03\pm 0.04$ ($^{\circ}$\,C yr$^{-1}$)  &   0.48     & Not significant   \\
 Temperature\tabnotemark{b}   & $7.29\pm 0.40$ ($^{\circ}$\,C)     & $0.02\pm 0.01$ ($^{\circ}$\,C yr$^{-1}$)  &   0.06     & Not significant   \\
 Relative Humidity            & $34.71\pm 1.33$ (\%)               & $0.05\pm 0.20$ (\% yr$^{-1}$)             &   0.82     & Not significant  \\
 Atmospheric pressure         & $728.88\pm 0.27$ (mb)              & $0.12\pm 0.05$ (mb yr$^{-1}$)             &   0.03     & $1.1\pm 0.3$\,mb  \\
 Precipitation                & $124\pm 62$ (mm)                   & $32\pm  9$     (mm yr$^{-1}$)             &   0.01     & $286\pm 63$\,mm  \\
                
   \bottomrule
  \tabnotetext{a}{\small Coefficients of the linear regression $a + bx$ ($x = 1, 2,...\,N$ years.).}
  \tabnotetext{b}{\small Using data from 1969--2019 period.}
  \tabnotetext{c}{\small P value of the slope coefficient of the linear regression.  }
  \end{tabular}
   \end{changemargin}
\end{table}

Additionally, we searched for a change in the number of cold days (daily mean temperature $\le0^{\circ}$\,C) 
or the number of warm days (mean temperature $\ge17^{\circ}$\,C). These two extreme temperatures occur less 
than 10\% of the time. We present these statistics in Table~\ref{tab:NumberTemp}.  Once more, the linear fits to 
these data is not statistically significant. 
 

\begin{table}[!ht]\centering
  \setlength{\tabnotewidth}{0.5\columnwidth}
  \tablecols{3}
  \setlength{\tabcolsep}{1.\tabcolsep}
    \caption{Number of warm and cold days (2009--2019).} \label{tab:NumberTemp}
 \begin{tabular}{lrr}
    \toprule
            Year     & \multicolumn{1}{c}{No. cold}   & \multicolumn{1}{c}{No. warm}   \\
   \cmidrule{2-3}
                     & \multicolumn{1}{c}{$T\le0^{\circ}$\,C} & \multicolumn{1}{c}{$T\ge17^{\circ}$\,C}  \\
   \cmidrule{1-3}
        2009 &      40 &     10   \\
        2010 &      39 &     15   \\
        2011 &      56 &     14   \\
        2012 &      40 &     11   \\
        2013 &      40 &     10   \\
        2014 &      19 &      9   \\
        2015 &      28 &     16   \\
        2016 &      32 &     20   \\
        2017 &      34 &      7   \\
        2018 &      28 &     18   \\
        2019 &      42 &     17   \\
   \bottomrule
  \end{tabular}
\end{table}

\subsection{Relative humidity}
\label{sec:hum}

\begin{figure}[!ht]\centering
  \includegraphics[width=\columnwidth]{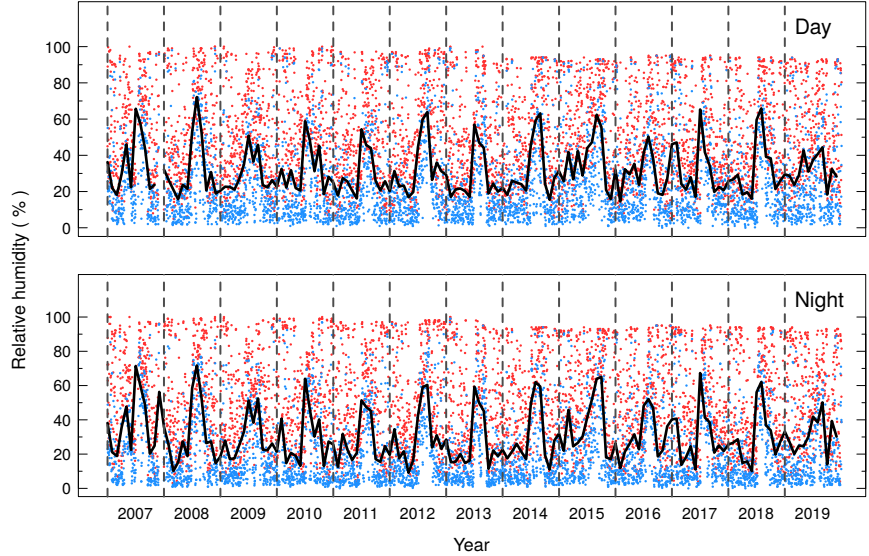}
  \caption{We present the thirteen year evolution (2007--2019) of the daily maximum (red) and minimum (blue) 
  relative humidity for daytime and nighttime data.  In each panel, the black line represents the median monthly 
  relative humidity, while the vertical dashed lines indicate the first day of each year.  The monthly median 
  relative humidity peaks in summer due to the North American monsoon.}
  \label{fig:humid}
\end{figure}

We present the daily maximum and minimum values of the relative humidity from 2007 to 2019 in 
Figure~\ref{fig:humid} for daytime and nighttime data.  In this figure, daily maximum and minimum relative 
humidity values and the monthly median relative humidity for each year are shown. The relative humidity 
covers a range from 0 to 100\,\% and shows a seasonal variation due to the North American monsoon 
\citep{1993JoC61665}, with summer months being the most humid on average.

\begin{figure*}[!ht]\centering
  \setlength\thisfigwidth{0.5\linewidth}
  \addtolength\thisfigwidth{-0.5cm}
  \makebox[\thisfigwidth][l]{\textbf{a}}%
  \hfill%
  \makebox[\thisfigwidth][l]{\textbf{b}}\\[-3ex]
  \parbox[t]{\linewidth}{%
     \vspace{0pt}
     \includegraphics[width=6cm,height=6cm]{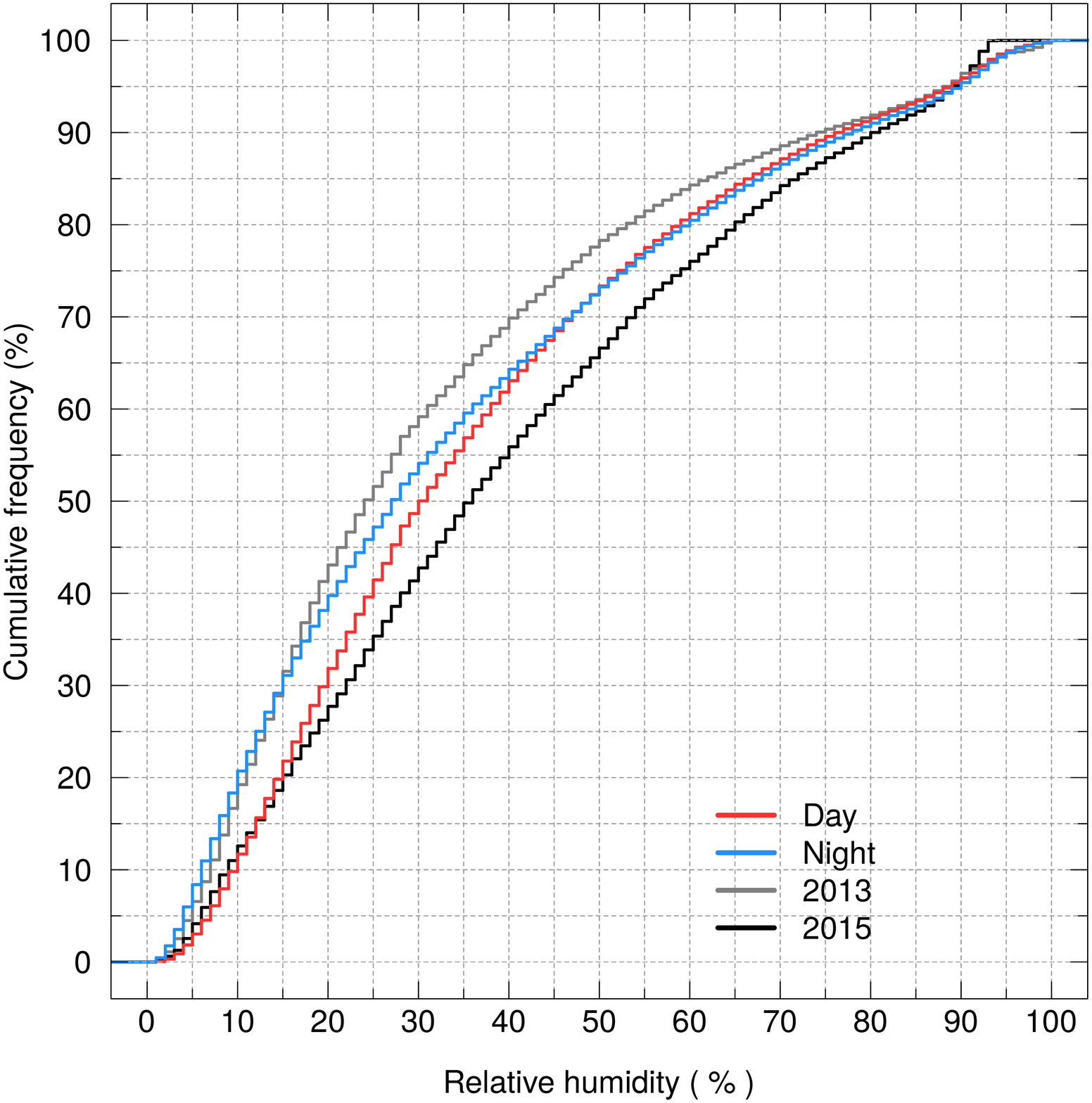}%
     \hfill%
     \includegraphics[width=6cm,height=6cm]{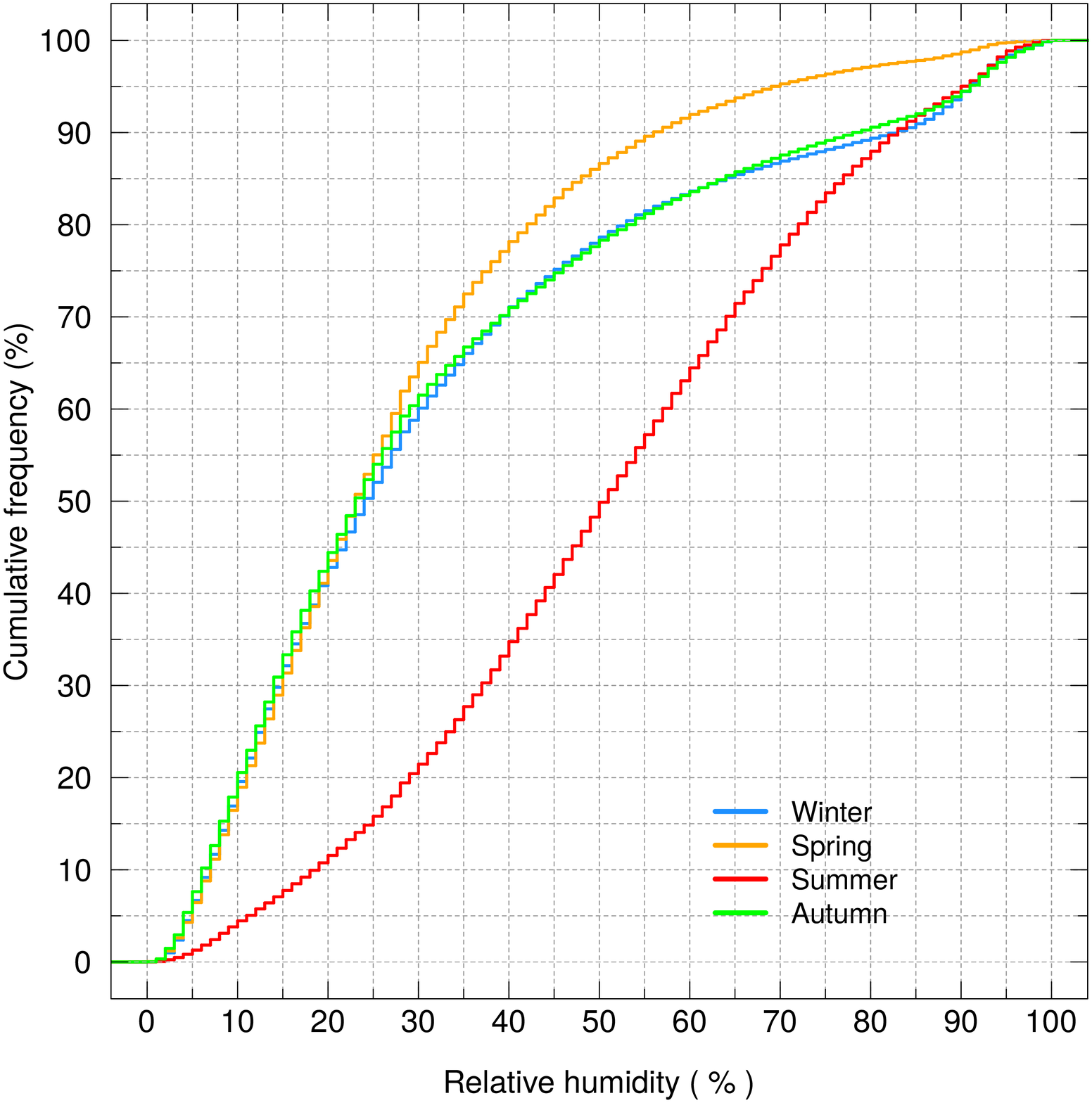}
     }
  \caption{We present (\textit{a}) the cumulative distribution of the daytime and nighttime relative humidity 
  obtained in the period 2007--2019, based upon the 5 minute averages. The blue line is for nighttime data, 
  red line is for daytime data, gray line is for 2013 (driest year) data and black line is for 2015 data 
  (most humid year), both data sets include day and nigth data. There is an excess of low humidity values 
  at night compared to during the day. In (\textit{b}) we present the cumulative distribution by season in 
  the same period. The blue line is for winter, orange line for spring, red line for summer and green line 
  for autumn.}
  \label{fig:humidD}
\end{figure*}

Figure~\ref{fig:humidD}a presents the cumulative distribution of the daytime and nighttime relative humidity 
based upon 5 minute averages.  The OAN-SPM is indeed a low humidity site, with a median relative humidity of 
30\% during the day and 27\% at night.  Though the distributions are very similar, a Wilcoxon-Mann-Whitney test 
confirms that they differ significantly (P value = $8.9\times10^{-9}$), with the median daytime relative 
humidity (30\,\%) being slightly greater than the median nighttime value (27\,\%). 

The drier conditions at night might seem counterintuitive, but they are mostly due the so called mountain 
thermal circulation. In their study of the hourly temperature and relative humidity variations at different altitudes 
at the Kilimanjaro, \citealp{2008AAAR402323}  found that at low altitudes the relative humidity increases during the 
night, while at high altitudes this increase is seen during the day (see their Figure 9b). According to these authors, 
the diurnal variation in humidity it is the outcome of strong upslope moisture transport during the day (via anabatic winds), 
counterbalanced by downslope transport and drying at night (via katabatic winds). This moisture is produced by the 
evapotranspiration process of the forest vegetation. At the OAN-SPM site, this effect has also been observed. In their 
study of characterization of Vallecitos site (a valley at an altitude of 2435\,m and $\sim$3.5\,km away from the 
OAN-SPM site), \citealp{2016PASP..128c5004T} found that at Vallesitos site the humidity is much higher at night 
that in the day, while for the OAN-SPM site it is slightly higher during the day (see  their Figure 12). These authors suggest 
that this might be the result of anabatic and katabatic winds (see their Figure 16). From Figure~\ref{fig:humidD}a, it can 
be seen that 75\% of the time, the relative humidity at night is lower than in the day, the rest of the time it is quite similar. 
The standard requirement for the use of the telescopes is that relative humidity be $\le$85\,\% (and that there 
be no condensation). At the OAN-SPM, such humid conditions ($>$85\,\% ) occur during only 6.5\,\% of daytime 
(300 hours per year) and 7.1\,\% nighttime (273 hours per year).

In Figure~\ref{fig:humidD}b we present the cumulative distribution by season. From this figure, 
it can be seen that summer is the most humid season with a median value of the relative humidity of 51\% and the 
driest is spring with a median value of 23\%. This is also shown in Table~\ref{tab:humid}, where we present 
monthly mean values of the relative humidity.  April to June (spring) is the driest period at OAN-SPM site 
while July to September (summer) is the most humid period (see Figure~\ref{fig:allparam}).

\begin{table}[!ht]\centering
  \setlength{\tabnotewidth}{0.5\columnwidth}
  \tablecols{4}
  \setlength{\tabcolsep}{1.5\tabcolsep}
  \caption{Monthly mean relative humidity} \label{tab:humid}
 \begin{tabular}{lrrr}
    \toprule
               & Day  & Night    &  $\mathrm{Day}+\mathrm{Night}$ \\
   \cmidrule{2-4}
    Month      & \multicolumn{3}{c}{(\%)} \\
   \cmidrule{1-4}
    January    & $36\pm 25$ & $36\pm 26$   & $36\pm 26$\\
    February   & $34\pm 27$ & $32\pm 26$   & $33\pm 27$\\
    March      & $31\pm 21$ & $27\pm 23$   & $29\pm 21$\\
    April      & $28\pm 18$ & $26\pm 20$   & $27\pm 18$\\
    May        & $31\pm 19$ & $27\pm 20$   & $29\pm 18$\\
    June       & $25\pm 14$ & $23\pm 16$   & $24\pm 15$\\
    July       & $52\pm 21$ & $52\pm 21$   & $52\pm 21$\\
    August     & $53\pm 21$ & $52\pm 21$   & $52\pm 20$\\
    September  & $48\pm 24$ & $47\pm 25$   & $47\pm 24$\\
    October    & $33\pm 22$ & $30\pm 25$   & $31\pm 23$\\
    November   & $31\pm 23$ & $29\pm 24$   & $30\pm 23$\\
    December   & $35\pm 28$ & $33\pm 26$   & $34\pm 27$\\

    \bottomrule
  \end{tabular}
\end{table}

In Table~\ref{tab:humidY}, we present the annual median relative humidity for the years 2007 to 2019.   
Considering day and night data, the driest year was 2013 and the most humid 2015, with relative humidities 
of 24\,\% and 35\,\%, respectively. This can be seen in Figure~\ref{fig:humidD}a, where we show the 2013
and 2015 cumulative distribution of the relative humidity.

\begin{table}[!ht]\centering
  \setlength{\tabnotewidth}{0.5\columnwidth}
  \tablecols{4}
  \setlength{\tabcolsep}{1.0\tabcolsep}
  \caption{Annual median relative humidity} \label{tab:humidY}
 \begin{tabular}{lccc}
    \toprule
         & Day & Night & $\mathrm{Day}+\mathrm{Night}$  \\
   \cmidrule{2-4}
    Year & \multicolumn{3}{c}{(\%)}\\
   \cmidrule{1-4}
   2007\tabnotemark{a}   & $33$   & $32$    & $34$   \\
   2008\tabnotemark{a}   & $28$   & $28$    & $28$   \\
   2009                  & $27$   & $27$    & $27$   \\
   2010                  & $31$   & $26$    & $29$   \\
   2011                  & $27$   & $27$    & $26$   \\
   2012                  & $30$   & $28$    & $28$   \\
   2013                  & $26$   & $22$    & $24$   \\
   2014                  & $27$   & $26$    & $27$   \\
   2015                  & $36$   & $32$    & $35$   \\
   2016                  & $30$   & $30$    & $29$   \\
   2017                  & $29$   & $28$    & $28$   \\
   2018                  & $28$   & $27$    & $27$   \\
   2019                  & $31$   & $29$    & $29$   \\

    \bottomrule
    \tabnotetext{a}{\small $< 95\%$ of data (see Table~\ref{tab:data}).}
  \end{tabular}
\end{table}

\begin{figure}[!ht]\centering
  \includegraphics[width=1.0\columnwidth]{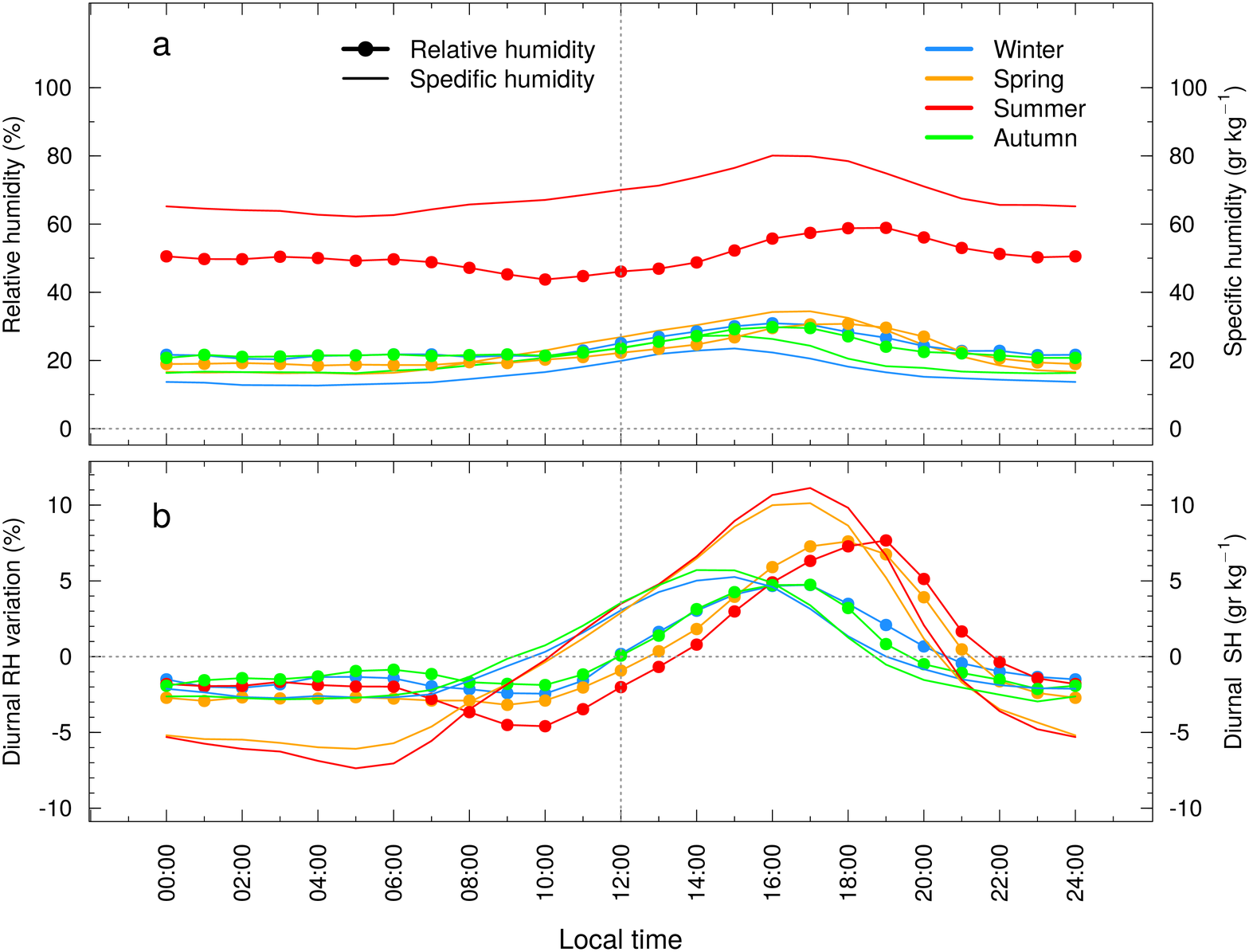}
  \caption{We present the relative and specific humidity (upper panel) and the diurnal relative and specific 
  humidity variation (bottom panel) by season.  Summer is the most humid season (50\% relative humidity), with 
  the others all being substantially lower (30\% relative humidity).  The diurnal variation of the humidity 
  is highest (and similar) in spring and summer while autumn and winter have a smaller diurnal variation.}
  \label{fig:humidH}
\end{figure}

Figure~\ref{fig:humidH} presents the hourly relative humidity and its variation by season.  The first panel 
(a) makes clear that summer is by far the most humid season.  There is little difference in relative humidity 
between the other seasons.  In the second panel (b), for each season, we subtract the mean relative humidity 
of the day from the relative humidity at a given hour during the day. At a given hour, values below the zero 
line indicate that the relative humidity is lower than the mean relative humidity of the day.  

In general, for all seasons, the relative humidity first decreases as the atmosphere warms up during the first 
hours after sunrise (morning warming), but then increases throughout the rest of the day (via anabatic winds). 
As the atmosphere cools at sunset, the relative humidity reaches its maximum value, before rapidly falling over 
the next few hours (via katabatic winds) and reaching a stable value shortly before midnight. Figure~\ref{fig:humidH}b 
indicates that the most extreme (and similar) diurnal variation in relative humidity occurs during the spring and summer seasons.

The relative humidity provides only a partial picture of the water content of the atmosphere at the OAN-SPM. 
Figure \ref{fig:humidH} also presents the specific humidity, the amount of water vapor (in grams) present in 
an air volume of 1 kg, which depends on air temperature, relative humidity and atmospheric pressure. 
The hourly specific humidity is plotted in absolute value in Figure~\ref{fig:humidH}a, while its 
diurnal variation is plotted in Figure~\ref{fig:humidH}b with continuous lines.

Based upon the specific humidity, there is a much greater amount of water vapor in the atmosphere in the 
summer than in the other seasons (3 times more; Figure~\ref{fig:humidH}a). This is not surprising given 
that both the temperature and the relative humidity are highest in summer, the two factors most affecting 
the specific humidity. The specific humidity is especially low in winter.

The diurnal variation in summer and spring are quite similar (see Figure~\ref{fig:humidH}b), with the only 
difference being a lower minimum at sunrise ($\sim$05:00 hours) for the summer. In detail, the specific 
humidity depends upon the relative humidity, the atmospheric pressure and the temperature, but temperature 
provokes variations of the largest absolute magnitude for the typical conditions at the OAN-SPM. Hence, 
the diurnal variation in the specific humidity is mostly driven by the diurnal variation in the temperature, 
though also scaled by the relative humidity. There is a greater diurnal temperature variation in spring than 
in summer (see Figure~\ref{fig:tempH}), but both a lower relative humidity (Table~\ref{tab:humid}) and a smaller 
amplitude in its diurnal variation. These factors conspire to produce a similar diurnal variations in the 
specific humidity in spring and summer. As for autumn and winter, their diurnal temperature variations are 
similar, as are their relative humidities, so they too have similar diurnal variations in the specific 
humidity.

To investigate whether the relative humidity varies with time, we fit the mean annual values. We find 
no significant variation as a function of time (see Table~\ref{tab:fitsVar}).


\subsection{Atmospheric pressure}
\label{sec:press}

\begin{figure}[!ht]\centering
  \includegraphics[width=1.0\columnwidth]{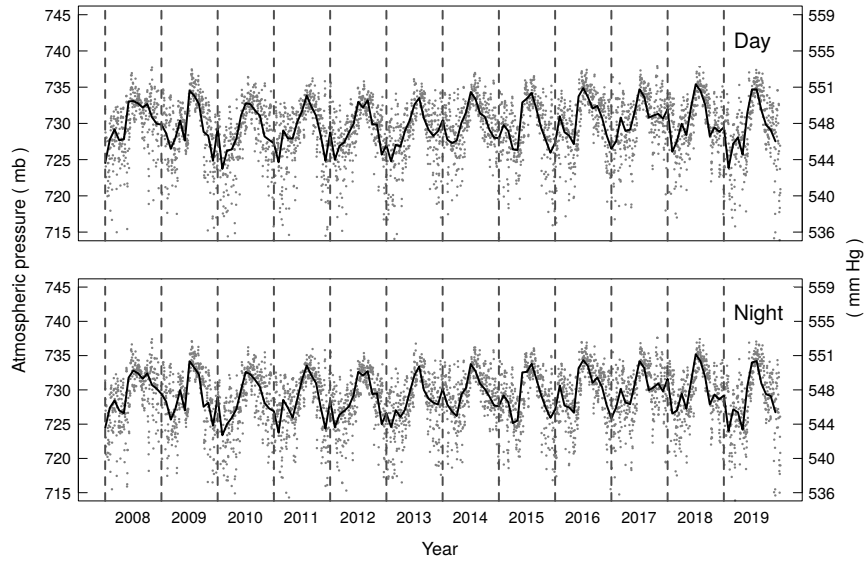}
  \caption{We present the variation of the daily mean atmospheric pressure for daytime and nighttime 
  data as a function of time (2008--2019). In each panel, the black curve represents the median 
  monthly atmospheric pressure and the vertical dashed lines indicate the first day of each year. 
  There is a clear seasonal variation, with the highest atmospheric pressure in summer and the 
  lowest in winter.  There is also greater variation in the atmospheric pressure in winter.}
  \label{fig:pressure}
\end{figure}

The daily median values of the atmospheric pressure are plotted in Figure~\ref{fig:pressure}. 
Note that we have no data from 2007, when the barometer was not working properly (Table \ref{tab:data}). 
There is a clear seasonal variation of the atmospheric pressure, with high values in summer and 
lower values in winter.  The atmospheric pressure has a median value of 730\,mb, but varies 
within the range of 713--737\,mb (534--552\,mm-Hg) during the entire period (see Figure~\ref{fig:allparam}). 
As for the air temperature, there is much more variation in the atmospheric pressure in winter 
than in summer.  

\begin{figure*}[!ht]\centering
  \setlength\thisfigwidth{0.5\linewidth}
  \addtolength\thisfigwidth{-0.5cm}
  \makebox[\thisfigwidth][l]{\textbf{a}}%
  \hfill%
  \makebox[\thisfigwidth][l]{\textbf{b}}\\[-3ex]
  \parbox[t]{\linewidth}{%
     \vspace{0pt}
     \includegraphics[width=6cm,height=6cm]{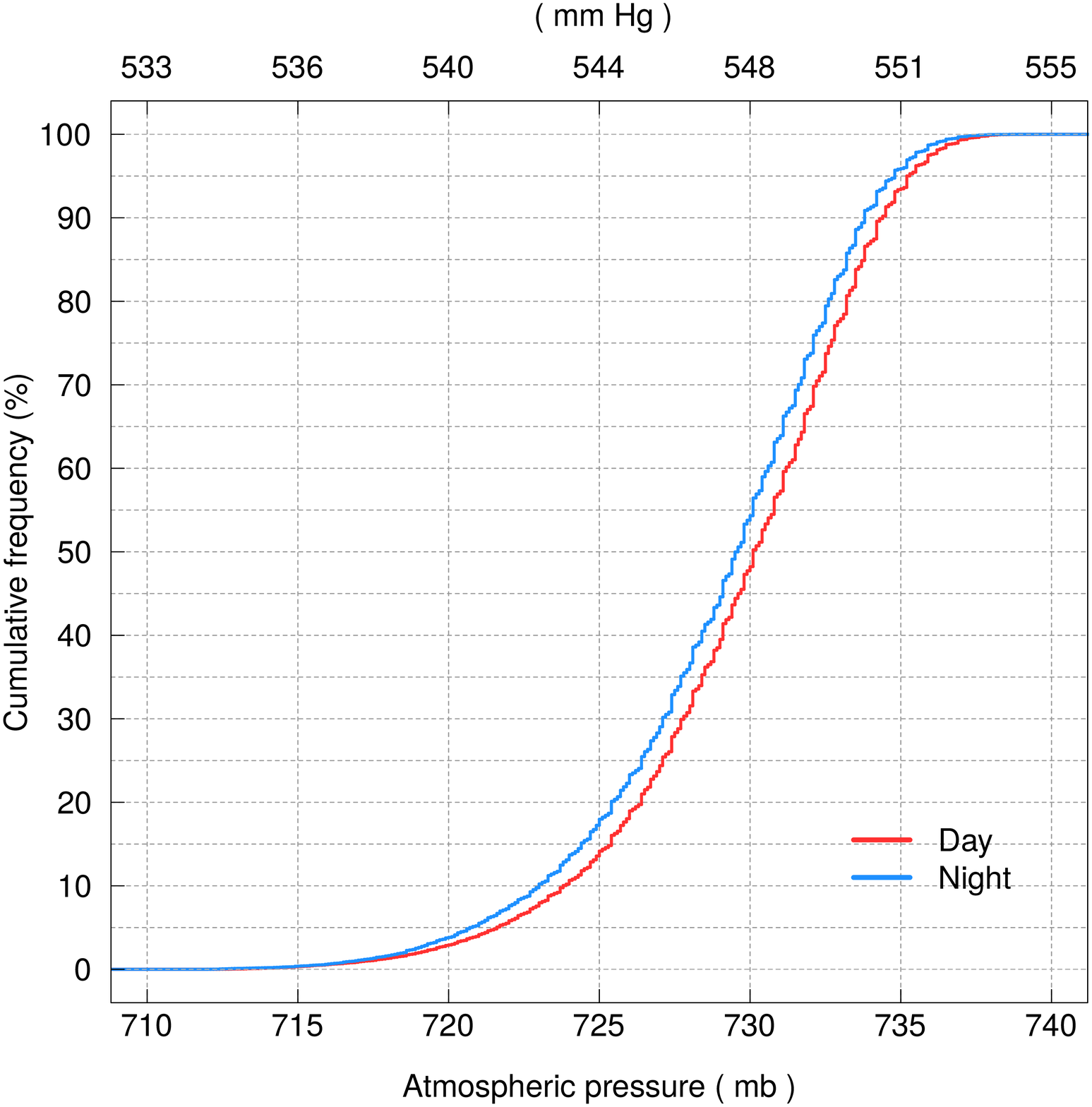}%
     \hfill%
     \includegraphics[width=6cm,height=6cm]{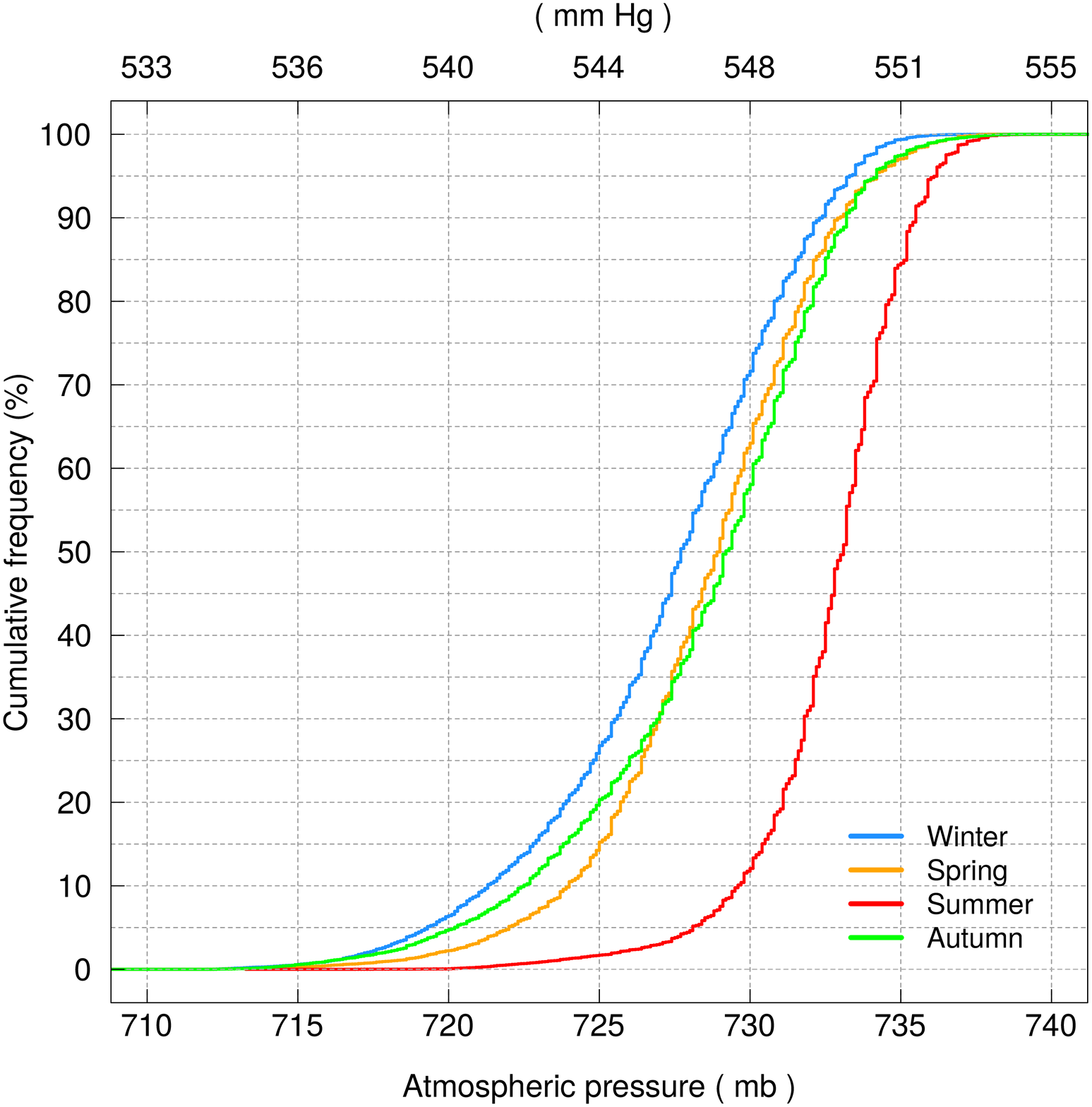}
     }
  \caption{We present (\textit{a}) the cumulative distribution of the daytime and nighttime atmospheric pressure 
  obtained in the  period 2008--2019, based upon the 5 minute averages. The blue line is for nighttime 
  data and the red line is for daytime data. There is a statistically-significant shift between the two 
  distributions, with the daytime values shifted to higher values.  In (\textit{b})  we present the cumulative 
  distribution by season in the same period. The blue line is for winter, orange line for spring, red line 
  for summer and green line for autumn.}
  \label{fig:pressureD}
\end{figure*}

In Figure~\ref{fig:pressureD}a, we present the cumulative  distribution of the daytime and nighttime 
atmospheric pressure obtained in the period 2008--2019, based upon the 5 minute averages. The median 
atmospheric pressure during the day is 730.1\,mb while at night it is 729.6\,mb.  Although the 
distributions for the day and night are similar, there is a systematic shift.  A Wilcoxon-Mann-Whitney 
test confirms that both distributions are statistically different ($P=8.4\times10^{-10}$ that the two 
distributions arise from the same parent distribution), with the median daytime atmospheric pressure 
value exceeding the nighttime median value by 0.5\,mb. As we show later, this difference 
is mainly due to the thermal atmospheric tides phenomenon.

In Figure~\ref{fig:pressureD}b, we present the cumulative distribution of atmospheric pressure by season. 
From this figure, it can be seen that the season with the highest atmospheric pressure 
is summer with a median value of 733.1\,mb. On the other hand, winter has the lowest atmospheric pressure, 
with a median value fo 727.7\,mb.

In Table~\ref{tab:pressure}, we present the monthly mean values of the atmospheric pressure. 
The seasonal variation seen in Figure \ref{fig:pressure} appears in the run of the values. During 
the summer months (July to September), the  atmospheric pressure is the highest ($\sim733$\,mb). 
This occurs because the OAN-SPM samples the mid-layers of the low atmosphere air.  In summer, the 
air at lowest altitudes is heated most and so expands, which raises a greater fraction of the lower 
atmosphere to greater heights and raises the air pressure at middle heights in the process. The 
rest of the year, the median atmospheric pressure is below $\sim730$\,mb (see Figure~\ref{fig:pressureD}b). 
As a comparison, at the sea level (e.g. Ensenada, B.C.), the atmospheric pressure ranges between 
$1002-1025$\,mb and the annual pattern is the inverse, with highest atmospheric pressure during the 
winter ($\sim1016$\,mb) and the lowest during the summer ($\sim1011$\,mb)\footnote{Data in the 2012-2019 
period from CICESE weather stations \texttt{http://redmar.cicese.mx/}}.

\begin{table}[!t]\centering
  \setlength{\tabnotewidth}{0.5\columnwidth}
  \tablecols{4}
  \setlength{\tabcolsep}{1\tabcolsep}
  \caption{Monthly median atmospheric pressure} \label{tab:pressure}
 \begin{tabular}{lrrr}
    \toprule
          & Day  & Night &  $\mathrm{Day}+\mathrm{Night}$ \\
   \cmidrule{2-4}
    Month & \multicolumn{3}{c}{(mb)} \\
   \cmidrule{1-4}
    January    & $728$ & $728$  & $728$   \\
    February   & $727$ & $727$  & $727$   \\
    March      & $728$ & $727$  & $728$   \\
    April      & $728$ & $727$  & $728$   \\
    May        & $728$ & $727$  & $728$   \\
    June       & $731$ & $730$  & $731$   \\
    July       & $734$ & $733$  & $734$   \\
    August     & $734$ & $733$  & $733$   \\
    September  & $732$ & $732$  & $732$   \\
    October    & $730$ & $730$  & $730$   \\
    November   & $729$ & $729$  & $729$   \\
    December   & $728$ & $727$  & $728$   \\

    \bottomrule
  \end{tabular}
\end{table}

\begin{figure}[!ht]\centering
  \includegraphics[width=1.0\columnwidth]{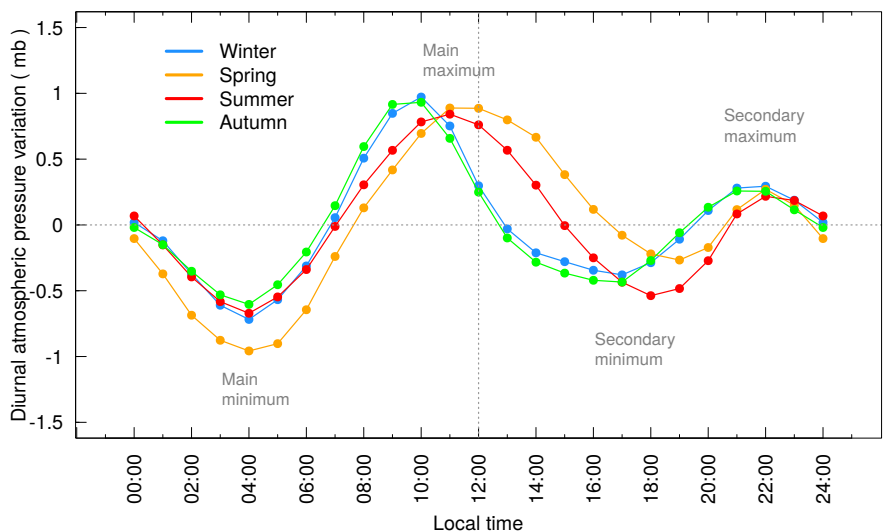}
  \caption{We present the diurnal and semidiurnal atmospheric pressure variation by season. To construct 
  these curves, we subtract the daily mean value from the hourly means and then average all of the data 
  by season and hour. Diurnal atmospheric pressure variation consists of the main maximum and secondary minimum, 
  while the semidiurnal variation consists of the main minimum and the secondary maximum.}
  \label{fig:pressureH}
\end{figure}

Figure~\ref{fig:pressureH} presents the variation of the atmospheric pressure throughout the day.  Here, 
we bin the data by season and subtract the daily mean value.  The atmospheric pressure varies through 
two cycles per day whose shape varies seasonally. These variations, also known as thermal atmospheric tides, 
are complex and not fully understood, but their main cause is the absorption of the solar radiation by the 
ozone in the stratosphere, aided by water vapor (\citealp{1961AdvGeo..7..105S,1970AtmospTides201,1973PAG193,1987JWS472}). 
In the upper atmosphere, the diurnal heating cycle gives rise to diurnal pressure waves, but the dynamic 
structure of the atmosphere causes the semidiurnal harmonic to be dominant (\citealp{1987JWS472,2011RMS..66..11L}). 
Hence, we observe two cycles per day.  

The diurnal pressure variation has a minimum and a maximum that occurrs at the same time each day in all 
seasons. In Figure~\ref{fig:pressureH}, the diurnal minimum (main minimum) occurs at 04:00\,hrs while the 
diurnal maximum (secondary maximum) occurs around 22:00\,hrs. On the other hand, the semidiurnal pressure 
variation varies seasonally.  Its maximum (main maximum) happens earlier in winter and autumn (10:00\,hrs) 
compared to spring and summer (12:00\,hrs) and, likewise, its minimum (secondary minimum) also occurs earlier 
in winter and autumn (16:00\,hrs) than in spring and summer (19:00\,hrs). Overall, the net effect of the atmospheric 
tides is a lower atmospheric pressure at night compared to the day. The atmospheric pressure variation 
during the night has a mean value of $-0.64$, while this value is $0.25$ during the day.

Studies indicate that the amplitude of atmospheric tides varies with latitude, from about 0.3\,mb in polar 
regions to 3.0\,mb in the tropics (\citealp{2011RMS..66..11L}).  The amplitude of the tides seen at the OAN-SPM 
fits this pattern, since we find an amplitude of $\sim2$\,mb. Also, there is evidence that the amplitude varies 
seasonally \citep{2011RMS..66..11L}. Our results support this finding, since the amplitude is approximately
1.7\,mb in winter, 1.8\,mb in spring, and 1.5\,mb for summer and autumn. 

Over the last ten years, the fit to our data shows that the annual mean atmospheric pressure increased 
by 1.1\,mb (see Table~\ref{tab:fitsVar}).

\subsection{Precipitation}
\label{sec:prec}

Previous studies of the climate at the OAN-SPM did not include precipitation.  These studies relied upon 
measurements obtained at weather stations in the region of the observatory, at distances up to around 150\,km 
\citep{2007RMxAC..31..113A} in order to understand the precipitation regime at the top of the mountain. 
Here, we include rain precipitation data recorded over the last thirteen years. As noted earlier, our data underestimates the precipitation from snowfall since the pluviometers are unreliable in sub-zero temperatures. 
Hence, our precipitation data should be treated as a lower limit to the true total precipitation. As 
mentioned in \S~\ref{sec:data}, we have no data for 2011.

\begin{figure}[!ht]\centering
  \includegraphics[width=1.0\columnwidth]{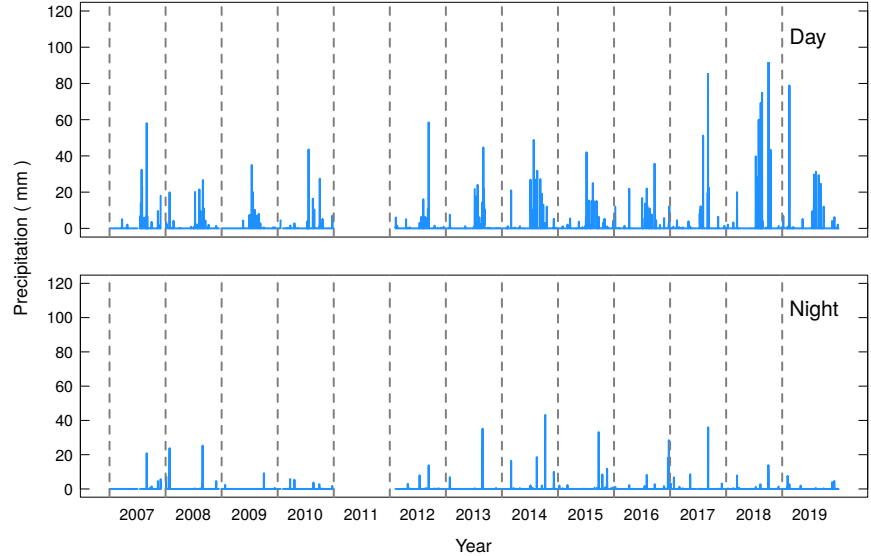}
  \caption{We present the records (2007--2019) of the daily accumulated precipitation for daytime and 
  nighttime data. The vertical dashed lines indicate the first day of each year.  No data was recorded for 
  2011. These data indicate that there is more precipitation during the day than at night.}
  \label{fig:precip}
\end{figure}

In Figure~\ref{fig:precip}, we show the daily accumulated precipitation for the period 2007 to 2019 for 
daytime and nighttime data.  It is clear that rain is more common during the day than at night. Overall, 
our records indicate that there is four times more rain during the day than at night.

Two factors affect the precipitation at the OAN-SPM. On the one hand, the rainy season in the northwest 
coastal region of Mexico occurs during the winter and it is primarily caused by the southern tails of 
North Pacific winter storms reaching Southern California and Northern Baja California.  On the other hand, 
in the area surrounding the north end of the Mar de Cortés (Gulf of California), heavy rainstorms occur 
during the summer, when the precipitation rises to a maximum in August due to the water vapor carried by 
the North American monsoon \citep{1997JCli...10.2600H}.  Since the Sierra San Pedro Mártir, where the 
OAN-SPM is located, occupies the highest reaches of the mountain range dividing these two regimes, 
it is affected by both.

In Table~\ref{tab:precip}, we present the monthly mean accumulated precipitation for daytime and nighttime, which 
is also plotted in Figure~\ref{fig:allparam}.  In Table~\ref{tab:PrecipMY}, we present the monthly 
accumulated precipitation for every year.  Both Tables \ref{tab:precip} and \ref{tab:PrecipMY}, as well 
as Figure \ref{fig:allparam}, demonstrate that the summer months dominate our precipitation data, though 
we should keep in mind that we are missing an important fraction of the winter precipitation.   

\begin{table}[!t]\centering
  \setlength{\tabnotewidth}{0.5\columnwidth}
  \tablecols{4}
  \setlength{\tabcolsep}{1.0\tabcolsep}
  \caption{Monthly mean accumulated precipitation} \label{tab:precip}
 \begin{tabular}{lrrr}
    \toprule
          & Day & Night & $\mathrm{Day}+\mathrm{Night}$  \\
   \cmidrule{2-4}
    Month & \multicolumn{3}{c}{(mm)}\\
   \cmidrule{1-4}
    January    & $6$   & $7$   & $13$\\
    February   & $11$  & $3$   & $14$\\
    March      & $4$   & $2$   & $5$\\
    April      & $3$   & $1$   & $5$\\
    May        & $2$   & $1$   & $2$\\
    June       & $3$   & $0$   & $3$\\
    July       & $76$  & $2$   & $77$\\
    August     & $80$  & $13$  & $92$\\
    September  & $44$  & $10$  & $54$\\
    October    & $20$  & $8$   & $27$\\
    November   & $5$   & $4$   & $9$\\
    December   & $5$   & $8$   & $12$\\

    \bottomrule
  \end{tabular}
\end{table}

\begin{table}[!t]\centering
  \begin{changemargin}{-3cm}{-2cm}
  \setlength{\tabnotewidth}{1.0\columnwidth}
  \tablecols{14}
  \setlength{\tabcolsep}{1.1\tabcolsep}
  \caption{Monthly accumulated precipitation (2007--2019)} \label{tab:PrecipMY}
 \begin{tabular}{lrrrrrrrrrrrrr}
    \toprule
            & 2007\tabnotemark{a} & 2008\tabnotemark{a} & 2009 & 2010 & 2011\tabnotemark{b} & 2012  & 2013  & 2014  & 2015  & 2016  & 2017  & 2018 & 2019 \\
   \cmidrule{2-14}
    Month   & \multicolumn{13}{c}{(mm)}  \\
   \cmidrule{1-14}
    Jan    & $0    $    & $72.7 $  & $3.2 $ & $4.4 $   & \nodata  & $0    $ & $30.3 $ & $0.1  $ & $4.1  $ & $14.2 $ & $14.6 $ & $2.4  $ & $8.5 $  \\
    Feb    & $0    $    & $4.8  $  & $0   $ & $0.4 $   & \nodata  & $13.1 $ & $0    $ & $36.1 $ & $0.1  $ & $0.1  $ & $7.7  $ & $3.9  $ & $100.0 $  \\
    Mar    & $7.5  $    & $0    $  & $0   $ & $5.6 $   & \nodata  & $0    $ & $0    $ & $3.4  $ & $14.5 $ & $1.5  $ & $0.8  $ & $25.7 $ & $1.0 $  \\
    Apr    & $2.7  $    & $0.30 $  & $0   $ & $9.0 $   & \nodata  & $8.4  $ & $0    $ & $1.6  $ & $0.1  $ & $26.5 $ & $3.9  $ & $0    $ & $2.1 $  \\
    May    & $0    $    & $0    $  & $5.8 $ & $0.2 $   & \nodata  & $0.9  $ & $1.2  $ & $0.1  $ & $3.9  $ & $0.1  $ & $8.8  $ & $0    $ & $5.2 $  \\
    Jun    & $0    $    & $0.9  $  & $12.2$ & $0   $   & \nodata  & $0    $ & $0    $ & $0    $ & $9.3  $ & $16.7 $ & $0    $ & $0.3  $ & $0   $  \\
    Jul    & $69.5 $    & $26.6 $  & $97.2$ & $56.6$   & \nodata  & $19.7 $ & $73.3 $ & $105.4$ & $78.4 $ & $67.0 $ & $55.5 $ & $188.0$ & $92.5 $  \\
    Aug    & $112.0$    & $129.6$  & $28.4$ & $44.8$   & \nodata  & $49.3 $ & $178.8$ & $85.7 $ & $68.5 $ & $46.2 $ & $70.8 $ & $197.9$ & $92.8 $  \\
    Sep    & $7.6  $    & $17.8 $  & $6.4 $ & $9.0 $   & \nodata  & $94.2 $ & $59.4 $ & $75.7 $ & $71.0 $ & $44.5 $ & $151.5$ & $15.8 $ & $89.9 $  \\
    Oct    & $7.2  $    & $2.0  $  & $9.8 $ & $38.6$   & \nodata  & $1.8  $ & $0.4  $ & $75.4 $ & $19.5 $ & $2.0  $ & $0    $ & $169.4$ & $0 $  \\
    Nov    & $39.5 $    & $7.6  $  & $0   $ & $0   $   & \nodata  & $1.2  $ & $0.9  $ & $0    $ & $23.5 $ & $8.8  $ & $6.6  $ & $0.3  $ & $14.7 $  \\
    Dec    & \nodata    & $0    $  & $2.2 $ & $10.2$   & \nodata  & $1.2  $ & $0.3  $ & $27.3 $ & $6.0  $ & $71.9 $ & $4.6  $ & $3.4  $ & $17.1 $  \\
    \midrule       
  Total    & $246.0$    & $262.3$  &$165.2$ & $178.8$  & \nodata  & $189.8$ & $344.6$ & $410.8$ & $298.9$ & $299.5$ & $324.8$ & $607.1$  & $423.8$    \\
  Day      & $224.5$    & $184.6$  &$152.2$ & $159.8$  & \nodata  & $159.2$ & $247.1$ & $294.6$ & $224.9$ & $222.3$ & $259.3$ & $573.5$  & $389.7$    \\
  Night    & $ 21.5$    & $ 77.7$  &$ 13.0$ & $ 19.0$  & \nodata  & $ 30.6$ & $ 97.5$ & $116.2$ & $ 74.0$ & $ 77.2$ & $ 65.5$ & $ 33.6$  & $ 34.1$    \\
                                                                                                                     
    \bottomrule
    \tabnotetext{a}{\small $< 95\%$ of data (see Table~\ref{tab:data}).}
    \tabnotetext{b}{\small Pluviometer not working.}
  \end{tabular}
 \end{changemargin}
\end{table}

\begin{figure}[!ht]\centering
  \includegraphics[width=1.0\columnwidth]{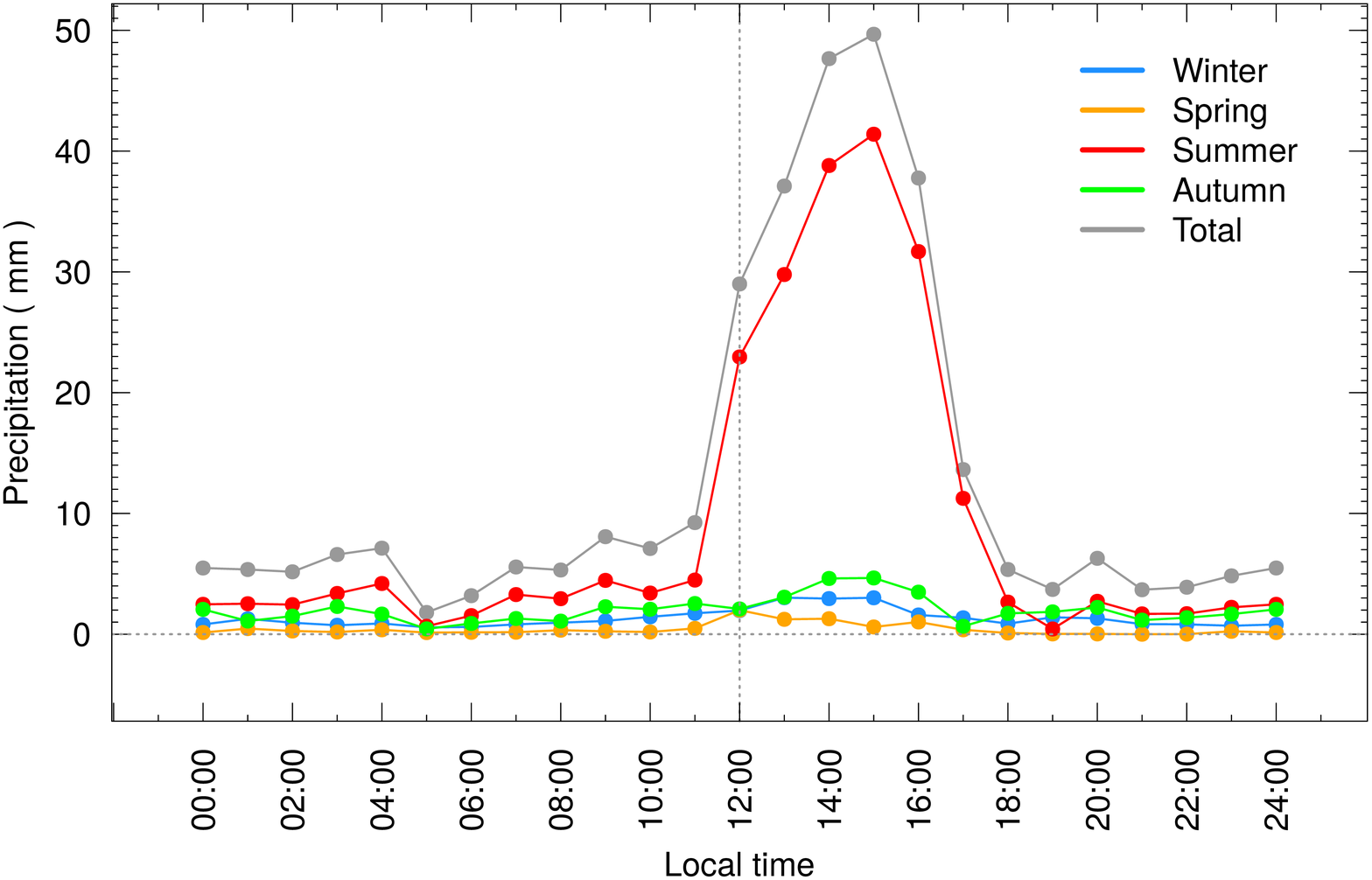}
  \caption{ We present the hourly accumulated precipitation by season as a function of time throughout the day.  
  Except in summer, there is no preferred time for precipitation.  In summer, most of the precipitation falls 
  as afternoon thunderstorms.}
  \label{fig:precipH}
\end{figure}

In Figure~\ref{fig:precipH}, we plot the accumulated precipitation according to the hour of the day for each 
season.  There is no noticeable trend, except for summer.  In summer, we find that the heaviest rains occur 
preferentially between 12:00 to 17:00\,hrs, which is to be expected since afternoon thunderstorms dominate 
the precipitation in summer.  In other seasons, the precipitation is similar for daytime and nighttime. Since 
the precipitation in other seasons is determined by larger-scale weather systems, this suggests that the snowfall 
in winter, largely absent in our data, is similar during the day and night.  Hence, we expect the true trend for 
winter in Figure~\ref{fig:precipH} to be a scaled version of the trend shown.

In the last three rows of Table~\ref{tab:PrecipMY}, we present the total ($\mathrm{day}+\mathrm{night}$), 
daytime and nighttime accumulated precipitation for 2007--2019. The annual mean rain 
precipitation at the OAN-SPM is $313\pm125$\,mm, with $\sim 70\%$ (220\,mm) falling between July and September. 
From Table~\ref{tab:PrecipMY}, the total is clearly very variable. 2018 had double the 
annual mean, in part due to abundant rains in October and more than usual during the summer, while 2009 was 
a year characterized by little precipitation. 

\citet{2007RMxAC..31..113A} report an annual rain precipitation of 449\,mm using a CONAGUA station (Comision 
Nacional del Agua) located at the National Park (13\,km southwest at an altitude of 2080\,m) during the period 
1977--2007, but only 173\,mm at the OAN-SPM. They report that 77\,\% of the precipitation at the National Park 
station occurs during winter time and the rest during spring and summer (see their Figure 2 and Table 3). 
They also include weather data for a station to the east of the Sierra San Pedro Mártir at a similar latitude 
as the other two stations.  There, precipitation almost only falls in summer, as a result of summer thunderstorms, 
and the yearly total amounts to 147\,mm.  Therefore, there is clearly a regime change in the precipitation in the 
vicinity of the OAN-SPM.  

At the OAN-SPM, we find at least 50\% more rainfall in summer (223\,mm; Table \ref{tab:precip}) than reported 
for the station further east, and double that reported by \citet{2007RMxAC..31..113A} for the National Park 
weather station.  Even if all of the winter precipitation recorded at the National Park station fell at the 
OAN-SPM, it would amount to only about half the total at the OAN-SPM.  If so, the total annual precipitation 
at the OAN-SPM is likely in the 400--450\,mm range.

Considering that the rainy days have, on average, a daily accumulated precipitation of 10\,mm, a heavy precipitation 
day could be one with $\ge 30$\, mm. In Table~\ref{tab:heavyRain} we present the dates with heavy accumulated 
precipitation in the 2007--2019 period. From this table it can be seen that heavy precipitation days are always 
during the summer days and in general characterized by high temperatures $\ge 10^{\circ}$\,C, high relative 
humidity ($>70$\,\%) and high atmospheric pressure ($\ge730$\,mb). On the other hand, heavy precipitation 
days where the relative humidity is low ($<70\%$) have the highest temperatures $>13^{\circ}$\,C. However, 
there were four dates with heavy precipitation, but low temperature ($<9^{\circ}$\,C) and low atmospheric 
pressure ($<728$\,mb). These might be the result of cold fronts, which produced the abundant precipitation recorded. 

Our findings are undoubtedly biased as a result of our partial winter precipitation records.  Clearly, our 
total annual precipitation must be an underestimate of the true value.  Correcting for this would increase the 
total precipitation in winter.  However, given the different nature of the storms, due to weather systems 
rather than thunderstorms, this winter precipitation is unlikely to be biased to daytime or nighttime. 
Hence, we expect the true winter precipitation in Figure \ref{fig:precipH} to be enhanced overall. 

\begin{table}[!ht]\centering
  \setlength{\tabnotewidth}{0.5\columnwidth}
  \tablecols{6}
  \setlength{\tabcolsep}{1\tabcolsep}
  \caption{Heavy precipitation\tabnotemark{a}} \label{tab:heavyRain}
  \begin{tabular}{rcrrrrr}
    \toprule
              &  Day/Night & Daily    & T      & RH      &  Atmosph. \\
              &            & precip.  & mean   & mean    &  pressure \\
   \cmidrule{2-6}
    Date      &             &  (mm)    & ($^{\circ}$\,C) & (\%)  & (mb) \\
   \cmidrule{1-6}
   Jul 29, 2007   & D      & $32$   & $14.4$  & $66$  &   --       \\
   Aug 31, 2007   & D      & $58$   & $12.9$  & $93$  &   --       \\
   Jul 15, 2009   & D      & $35$   & $16.4$  & $58$  &   $737$  \\
   Jul 20, 2010   & D      & $44$   & $16.4$  & $67$  &   $733$  \\
   Sep 9,  2012   & D      & $59$   & $10.7$  & $98$  &   $731$  \\
   Aug 26, 2013   & N      & $35$   & $10.0$  & $95$  &   $730$  \\
   Aug 31, 2013   & D      & $45$   & $13.1$  & $67$  &   $734$  \\
   Jul 25, 2014   & D      & $49$   & $15.8$  & $62$  &   $734$  \\
   Aug 17, 2014   & D      & $32$   & $14.3$  & $70$  &   $735$  \\
   Oct  8, 2014   & N      & $43$   & $8.0$   & $91$  &   $728$  \\
   Jul  4, 2015   & D      & $42$   & $14.9$  & $61$  &   $732$  \\
   Sep 21, 2015   & N      & $33$   & $9.7$   & $93$  &   $729$  \\
   Sep 19, 2016   & D      & $36$   & $10.1$  & $94$  &   $734$  \\
   Aug  1, 2017   & D      & $51$   & $13.1$  & $85$  &   $734$  \\
   Sep  2, 2017   & D      & $85$   & $10.8$  & $94$  &   $729$  \\
   Sep  3, 2017   & N      & $36$   & $10.1$  & $95$  &   $730$  \\
   Jul 10, 2018   & D      & $40$   & $11.6$  & $88$  &   $734$  \\
   Jul 29, 2018   & D      & $60$   & $13.7$  & $85$  &   $735$  \\
   Aug 11, 2018   & D      & $69$   & $12.6$  & $82$  &   $734$  \\
   Aug 20, 2018   & D      & $75$   & $15.3$  & $74$  &   $735$  \\
   Oct  1, 2018   & D      & $117$  & $8.4$   & $94$  &   $728$  \\
   Oct 15, 2018   & D      & $43$   & $2.1$   & $67$  &   $727$  \\
   Feb 14, 2019   & D      & $79$   & $4.5$   & $94$  &   $723$  \\
   Ago  6, 2019   & D      & $31$   & $15.3$  & $57$  &   $735$  \\
    \bottomrule

    \tabnotetext{a}{\small Acummulated precipitation  ($\ge 30$\,mm)}
   \end{tabular}
\end{table}

\begin{figure}[!ht]\centering
  \includegraphics[width=1.0\columnwidth]{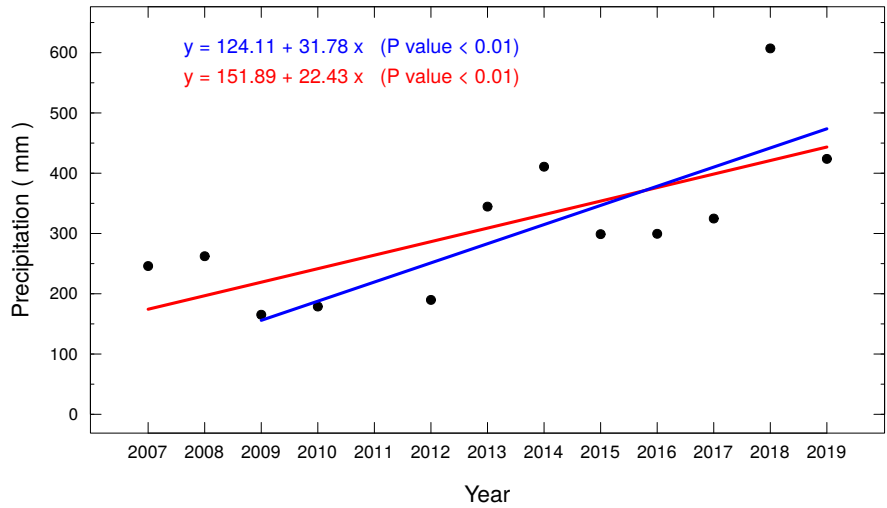}
  \caption{ We present the linear fits to annual precipitation data for the 2009--2019 period 
  (blue line) and 2007--2019 period (red line). Both linear fits have confidence level of 99\%.}
  \label{fig:precipInc}
\end{figure}

The yearly totals in Table \ref{tab:PrecipMY} show a clear increase in precipitation in recent years 
compared to the years up to 2012. Alternatively, if we fit these data as a linear trend, we find an 
annual increase in precipitation, on average, of 30\,mm more precipitation per year for the 2009--2019 
period (99\% confidence level). The values and uncertainties of the coefficients of the linear fit are 
shown in Table~\ref{tab:fitsVar}. In Figure~\ref{fig:precipInc} we present the linear fit to our 
data in the 2009-2019 period (blue line). In addition, we have included the linear fit for the 2007-2019 
period (2007 and 2008 are years with $<$95\% of data; see Table~\ref{tab:data}) to verify if this increase 
in precipitation over the years is also maintained in this period. We find that this trend continues, 
although it is slightly lower with 20\,mm per year (99\% confidence level). \citet{2007RMxAC..31..113A} 
present a weather model (see their Figure 7) that predicted that precipitation would increase from 2007, 
equalling its long-term average around 2015 and increasing to a maximum between 2020 and 2035.  Hence, the 
increase in total annual precipitation that we find supports their scenario.

Studies of global warming have shown that regions with the largest increases in heavy precipitation 
events include several high-latitude regions, mountainous regions, eastern Asia and eastern North 
America (\citealp{IPCC2018}). We count the number of days per year with measurable daily precipitation 
($P\ge1$\,mm) and the number of days with daily heavy precipitation ($P\ge10$\,mm) for the 2009--2019 
period (only years with $\ge$95\% data) and present this information in Table~\ref{tab:NumberPrecip}. 
We do not find any  significant variation in the number of days with precipitation or heavy precipitation, 
but rather a correlation with the total precipitation.

\begin{table}[!ht]\centering
  \setlength{\tabnotewidth}{0.5\columnwidth}
  \tablecols{3}
  \setlength{\tabcolsep}{1.\tabcolsep}
    \caption{Number of days with precipitation (2009--2019).} \label{tab:NumberPrecip}
 \begin{tabular}{lcc}
    \toprule
            Year     & \multicolumn{1}{c}{Number of days}   & \multicolumn{1}{c}{Number of days}   \\
   \cmidrule{2-3}
                     & \multicolumn{1}{c}{$P\ge$1\,mm} & \multicolumn{1}{c}{$P\ge$10\,mm}  \\
   \cmidrule{1-3}
        2009 &      20       &    5        \\
        2010 &      23       &    5        \\
        2011 &      \nodata  &   \nodata   \\
        2012 &      23       &    4        \\
        2013 &      25       &   13        \\
        2014 &      31       &   14        \\
        2015 &      35       &    9        \\
        2016 &      31       &   11        \\
        2017 &      28       &    7        \\
        2018 &      25       &   13        \\
        2019 &      35       &   12        \\
   \bottomrule                            
  \end{tabular}
\end{table}

\subsection{Wind speed and direction}
\label{sec:wind}

The DI and VWT weather stations deliver two wind speeds: the sustained wind speed and the wind gust speed. 
For a description of how these two parameters are determined, we refer the reader to \S\ref{sec:data}.

The DI and VWT stations were installed at different heights in the three sites, so we have normalized all 
wind speeds (both sustained and gusts) to a common 7 m height using Equation (1) of \citet{2010RMxAA..46...89B} 
(hereinafter referred to as BN10). Their fit to the median value of the wind speed as a function of height, $h$, is 

\begin{equation}
  \label{eq:one}
  V(h)/V(7) = 1.0103 + 0.4293\, \ln (h/7)\ 
\end{equation}

\begin{figure}[!ht]\centering
  \includegraphics[width=1.0\columnwidth]{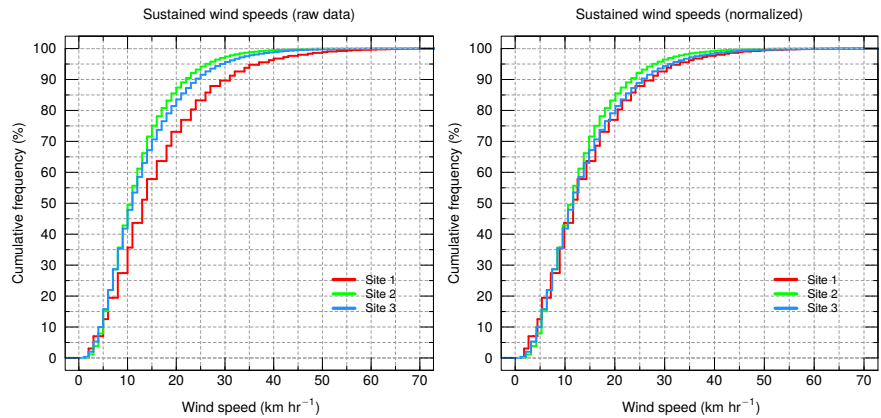}
  \caption{We present the cumulative distributions of the sustained wind speeds for the three sites before 
  (left panel) and after (right panel) applaying the normalization from \citet{2010RMxAA..46...89B}. These graphs 
  include data for the period 2007--2019, based upon the 5 minute average values.}
  \label{fig:windnorm}
\end{figure}

While Eq. 1 was derived for another nearby site, a relation of this form is expected generally (e.g., 
\citealp{1988SpringerS}). In  Figure~\ref{fig:windnorm}, we present the cumulative distributions of the 
sustained wind speed for the three sites before and after applying the normalization from BN10. 
Clearly, Eq. 1 is reasonably effective at matching the three wind speed distributions from the three sites, 
especially since our focus is upon the general conditions at the OAN-SPM.  The small residual differences are 
presumably due to differences in the vertical variation of the wind speed at the three sites relative to that 
at the site studied by BN10. It is likely that the smaller differences between sites 1 and 3 reflect that 
these sites, like that studied by BN10, are on the south side of the ridge whereas site 2 is on the north side. 
Since the strongest winds come predominantly from southerly directions (e.g., Figure~\ref{fig:winddir2}), site 2 
should be somewhat more protected by the local terrain than are sites 1 and 3. This may explain why the wind 
speed distribution for site 2 has a slightly lower proportion of the highest wind speeds.

As noted in \S\ref{sec:data}, the distribution of wind directions for site 2 seems to be rotated by 
45$^{\circ}$ when compared to sites 1 and 3, probably due to an error in the orientation of the sonic 
anemometer at site 2. Alternatively, the orography or buildings around site 2 may be to blame.  On the other 
hand, the cumulative distribution for the wind speed does not appear anomalous. Given the discrepancy 
with its wind directions, we discard the wind direction data gathered at site 2 from this study. 
Hence, when only the wind speed is needed, we include the data from site 2.  When both the wind speed 
and direction are needed, we discard the data from site 2 because its wind direction parameter it is 
not reliable.  

\begin{figure}[!ht]\centering
  \includegraphics[width=1.0\columnwidth]{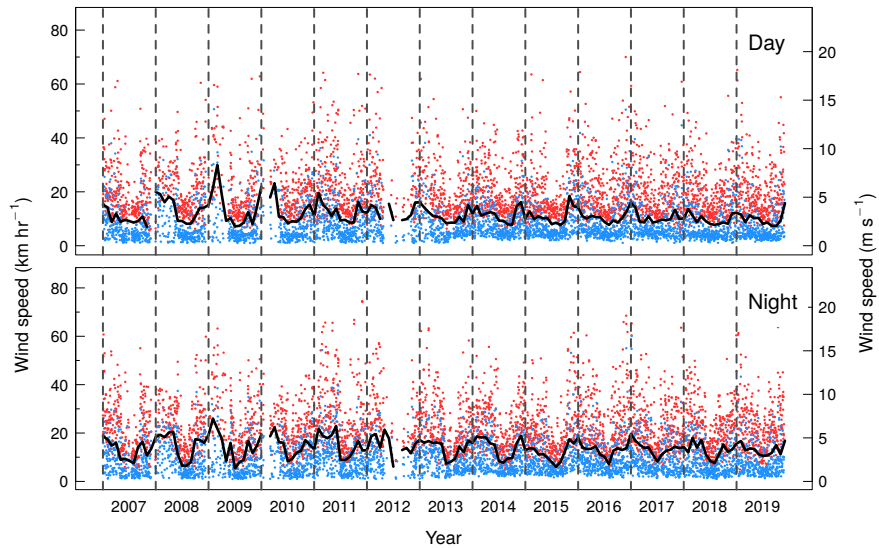}
  \caption{We present the daily maximum (red) and minimum (blue) sustained wind speeds for daytime and 
  nighttime data for the thirteen year time span of our data (2007--2019). In each panel the black line 
  represents the   median monthly wind speed and the vertical dashed lines indicate the first day of 
  each year. These graphs include data from sites 1, 2 and 3.}
  \label{fig:wind}
\end{figure}

The daily maximum and minimum values of the sustained wind speed are plotted in Figure~\ref{fig:wind} 
for daytime and nighttime. There are gaps in the data for the years 2007, 2008, 2009 and 2010 winters 
because the DI wind vane froze, and in 2012 the anemometer was broken (see Table ~\ref{tab:data}). From 
Figure~\ref{fig:wind}, it is also clear that the sustained wind speed is lowest in summer.

\begin{figure*}[!ht]\centering
  \setlength\thisfigwidth{0.5\linewidth}
  \addtolength\thisfigwidth{-0.5cm}
  \makebox[\thisfigwidth][l]{\textbf{a}}%
  \hfill%
  \makebox[\thisfigwidth][l]{\textbf{b}}\\[-3ex]
  \parbox[t]{\linewidth}{%
     \vspace{0pt}
     \includegraphics[width=6cm,height=6cm]{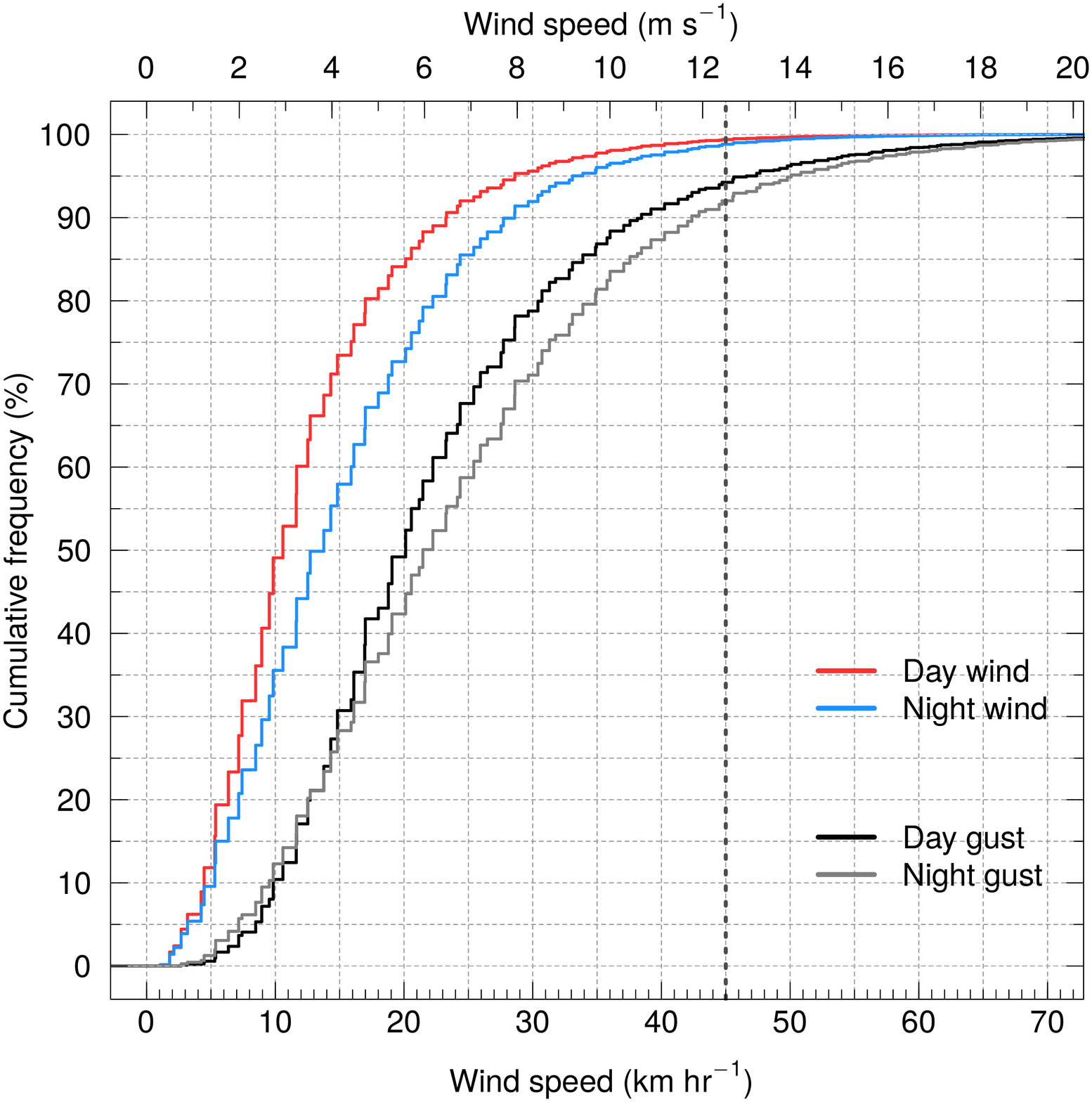}%
     \hfill%
     \includegraphics[width=6cm,height=6cm]{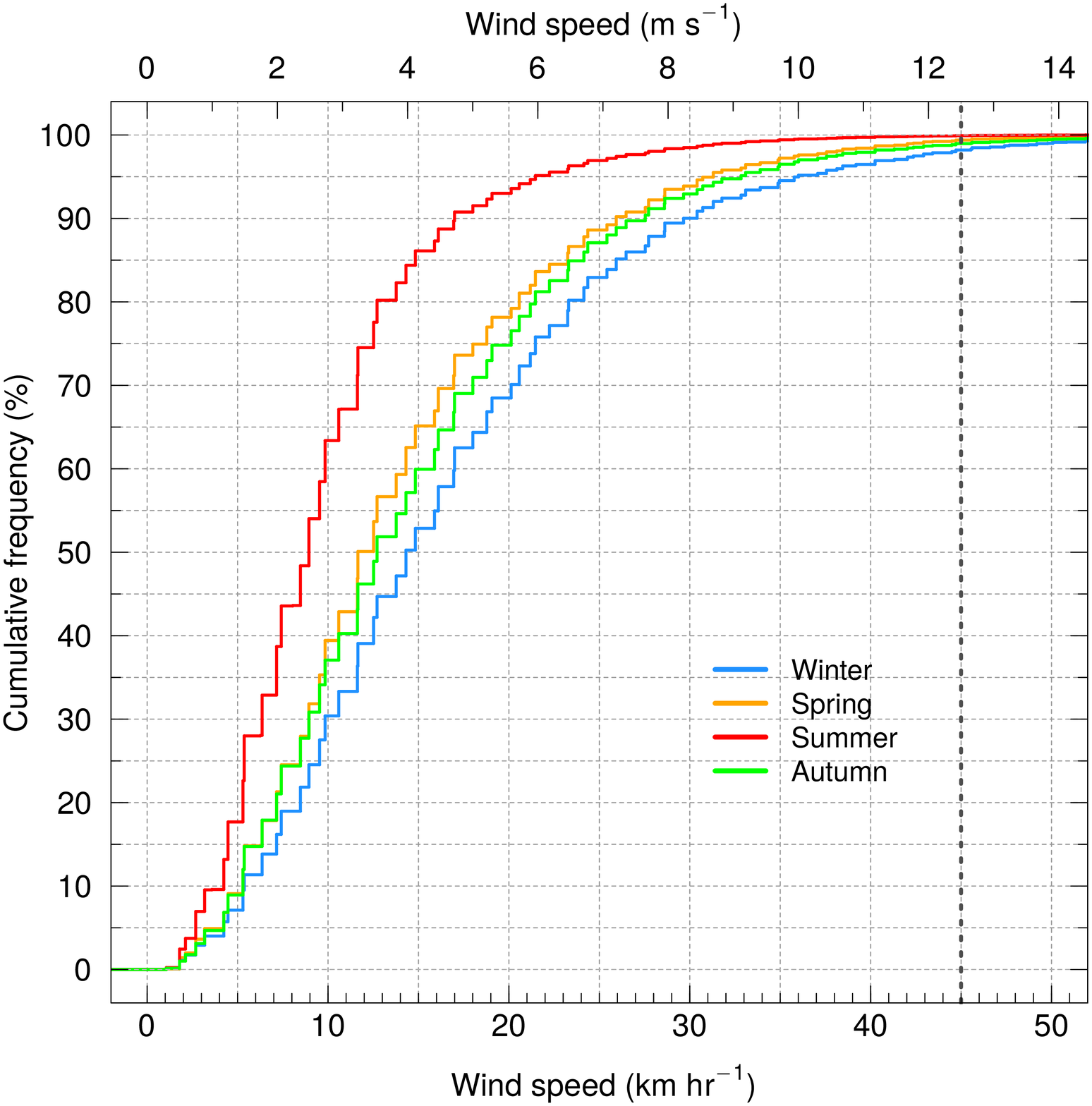}
     }
  \caption{We present (\textit{a}) the cumulative distribution of the daytime and nighttime sustained wind 
  speed and wind gusts for the period 2007--2019, based upon the 5 minute average and maximum 
  values, respectively. The blue line is for nighttime and the red line is for daytime sustained wind data. 
  The gray line is for nighttime and the black line is for daytime gusts wind data.  In (\textit{b}) we 
  present the cumulative distribution of the sustained wind speed by season in the same period. The blue line 
  is for winter, orange line for spring, red line for summer and green line   for autumn. In both graphs, 
  the  dashed vertical line indicates the sustained wind speed limit of 45\,km hr$^{-1}$ at which the domes 
  of the OAN-SPM telescopes are closed.  These graphs include data from sites 1, 2 and 3.}  
  \label{fig:windD}
\end{figure*}

In Figure~\ref{fig:windD}a and Table~\ref{tab:windDist}, we present the cumulative distribution of 
sustained wind and wind gust speeds for daytime and nighttime data. From this figure, it is clear that 
the wind speeds, both sustained and gusts, are stronger at night. The median sustained wind speed for 
daytime is 10.6\,km hr$^{-1}$ (2.9\,m s$^{-1}$) and at nighttime 12.9\,km hr$^{-1}$ (3.6\,m s$^{-1}$), 
while the median wind gust speeds are 20.1\,km hr$^{-1}$ (5.6\,m s$^{-1}$) and 21.5\,km hr$^{-1}$ 
($\sim$6.0\,m s$^{-1}$) for daytime and nighttime, respectively.

The cumulative distributions for the speed of the daytime and nighttime sustained winds differ greatly 
in shape. In Figure~\ref{fig:windD}a, we also indicate the speed limit of 45\,km hr$^{-1}$ (12.5\,m s$^{-1}$) 
for the sustained wind at which the OAN-SPM's telescope domes are closed.  The fraction of time with the 
sustained wind speed above this limit is $1.2\%$ during nighttime or 51.1 hours per year 
(see Table~\ref{tab:windDist}).  

In Figure~\ref{fig:windD}b and Table~\ref{tab:windDist}, we present the cumulative distribution of sustained 
wind speed by season. From Figure~\ref{fig:windD}b, it can be seen that the season with the strongest winds 
is winter, with a median value of 14.3\,km hr$^{-1}$ (4.0\,m s$^{-1}$), while for the summer season 
this value is 8.9\,km hr$^{-1}$ (2.5\,m s$^{-1}$). Spring and autumn have median values of 
11.7\,km hr$^{-1}$ (3.3\,m s$^{-1}$) and 12.7\,km hr$^{-1}$ ($\sim$3.5\,m s$^{-1}$), respectively.

\begin{figure}[!ht]\centering
  \includegraphics[width=0.7\columnwidth]{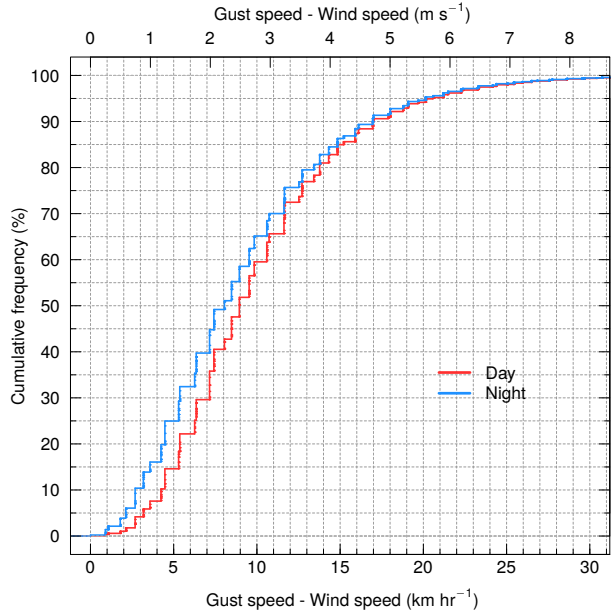}
  \caption{We present the cumulative distribution of the daytime and nighttime difference between wind gust
  and sustained wind speeds for the period 2007--2019, based   upon the 5 minute average and maximum values, 
  respectively. The blue line is for nighttime and the red line is   for daytime data. Values of the x-axis 
  higher than 30\,km hr$^{-1}$ represent less than 1\% of the cases. This graph includes data  from 
  sites 1, 2 and 3.}
  \label{fig:windDDiff}
\end{figure}

In Figure~\ref{fig:windDDiff},  we show the cumulative distribution of the difference between the wind  
gusts and sustained wind speeds. When compared to the sustained wind and wind gust speeds, we find that 
50\% of the time wind gusts are stronger than sustained winds by 8.1(nighttime) and 8.9\,km hr$^{-1}$(daytime) 
(2.3 and 2.5\,m s$^{-1}$). The difference between the wind gust and the sustained wind speeds rarely exceeds 
20\,km hr$^{-1}$ (5.6\,m s$^{-1}$), which occurs less than 6\% of the time.

\begin{table}[!t]\centering
  \begin{changemargin}{-3.5cm}{-2cm}
  \setlength{\tabnotewidth}{0.5\columnwidth}
  \tablecols{9}
  \setlength{\tabcolsep}{2\tabcolsep}
  \caption{Distribution of sustained wind and wind gust speeds} \label{tab:windDist}
 \begin{tabular}{crrrrrrrr}
    \toprule
    Speed           & \multicolumn{2}{c}{Wind} &  \multicolumn{2}{c}{Gusts} & \multicolumn{4}{c}{Wind} \\
   \cmidrule{2-9}
  (km hr$^{-1}$)    & Day   & Night            & Day       & Night           & Winter & Spring & Summer & Autumn \\
   \cmidrule{1-9}
  5     & 0.1183  & 0.0958  & 0.0058  & 0.0126  & 0.0711  & 0.0909  & 0.1768  & 0.0893  \\
  10    & 0.4909  & 0.3555  & 0.1041  & 0.1229  & 0.3039  & 0.3943  & 0.6338  & 0.3707  \\
  15    & 0.7345  & 0.5795  & 0.3071  & 0.2830  & 0.5287  & 0.6514  & 0.8611  & 0.5994  \\
  20    & 0.8409  & 0.7268  & 0.4920  & 0.4234  & 0.6848  & 0.7816  & 0.9301  & 0.7480  \\
  25    & 0.9202  & 0.8551  & 0.6765  & 0.5873  & 0.8293  & 0.8861  & 0.9693  & 0.8709  \\
  30    & 0.9559  & 0.9194  & 0.7876  & 0.7106  & 0.9001  & 0.9388  & 0.9848  & 0.9294  \\
  35    & 0.9777  & 0.9603  & 0.8685  & 0.8139  & 0.9453  & 0.9723  & 0.9942  & 0.9650  \\
  40    & 0.9870  & 0.9754  & 0.9104  & 0.8734  & 0.9647  & 0.9841  & 0.9972  & 0.9792  \\
  45    & 0.9939  & 0.9883  & 0.9425  & 0.9202  & 0.9819  & 0.9938  & 0.9991  & 0.9897  \\
  50    & 0.9969  & 0.9939  & 0.9621  & 0.9488  & 0.9900  & 0.9974  & 0.9998  & 0.9944  \\
  55    & 0.9986  & 0.9972  & 0.9754  & 0.9671  & 0.9949  & 0.9992  & 1       & 0.9974  \\
  60    & 0.9993  & 0.9985  & 0.9843  & 0.9789  & 0.9974  & 0.9996  & \nodata & 0.9986  \\
  65    & 0.9998  & 0.9995  & 0.9909  & 0.9872  & 0.9992  & 0.9999  & \nodata & 0.9994  \\
  70    & 0.9999  & 0.9998  & 0.9946  & 0.9924  & 0.9997  & 1       & \nodata & 0.9998  \\
  75    & 1       & 0.9999  & 0.9965  & 0.9951  & 0.9998  & \nodata & \nodata & 0.9999  \\
  80    & \nodata & 1       & 0.9980  & 0.9971  & 0.9999  & \nodata & \nodata & 1  \\
  85    & \nodata & \nodata & 0.9990  & 0.9984  & 1       & \nodata & \nodata & \nodata \\
  90    & \nodata & \nodata & 0.9994  & 0.9990  & \nodata & \nodata & \nodata & \nodata \\
  95    & \nodata & \nodata & 0.9997  & 0.9995  & \nodata & \nodata & \nodata & \nodata \\
  100   & \nodata & \nodata & 0.9998  & 0.9997  & \nodata & \nodata & \nodata & \nodata \\
  105   & \nodata & \nodata & 0.9999  & 0.9999  & \nodata & \nodata & \nodata & \nodata \\
  110   & \nodata & \nodata & 0.9999  & 0.9999  & \nodata & \nodata & \nodata & \nodata \\
  115   & \nodata & \nodata & 1       & 1       & \nodata & \nodata & \nodata & \nodata \\

    \bottomrule
  \end{tabular}
  \end{changemargin}
\end{table}

\begin{figure}[!ht]\centering
  \includegraphics[width=0.7\columnwidth]{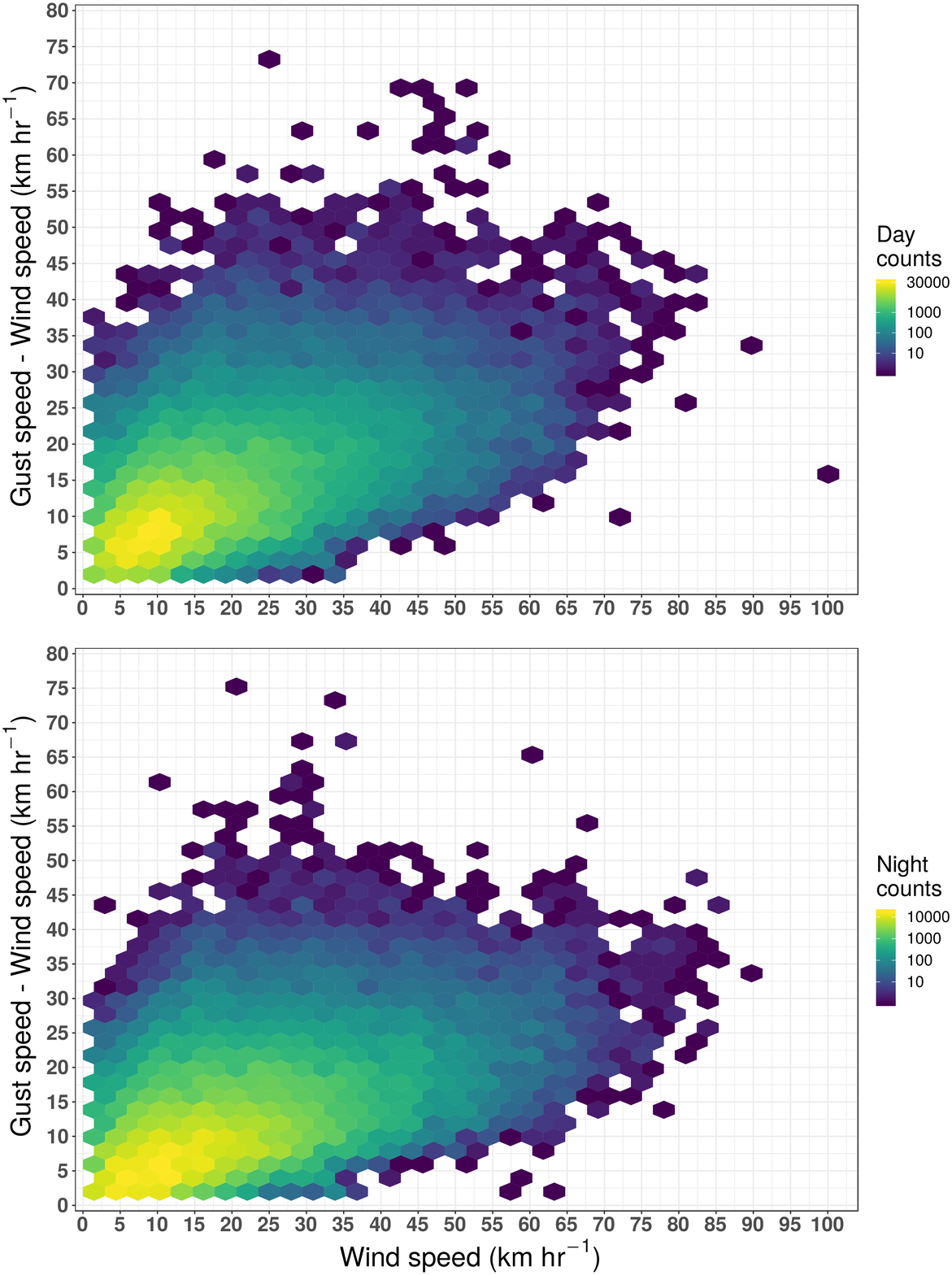}
  \caption{We present hexagonal bin plots of the daytime (upper panel) and nighttime (bottom panel) 
  difference between wind gust and sustained wind speeds as a function of sustained wind speed  
  based upon the 5 minute average (sustained wind) and maximum (wind gust) values for the period 2007--2019. 
  These graphs include data from sites 1, 2 and 3.}
  \label{fig:windDiff}
\end{figure}

\begin{figure}[!ht]\centering
  \setlength\thisfigwidth{0.5\linewidth}
  \addtolength\thisfigwidth{-0.5cm}
  \makebox[\thisfigwidth][l]{\textbf{a}}%
  \hfill%
  \makebox[\thisfigwidth][l]{\textbf{b}}\\[-3ex]
  \parbox[t]{\linewidth}{%
     \vspace{0pt}
     \includegraphics[width=6cm,height=6cm]{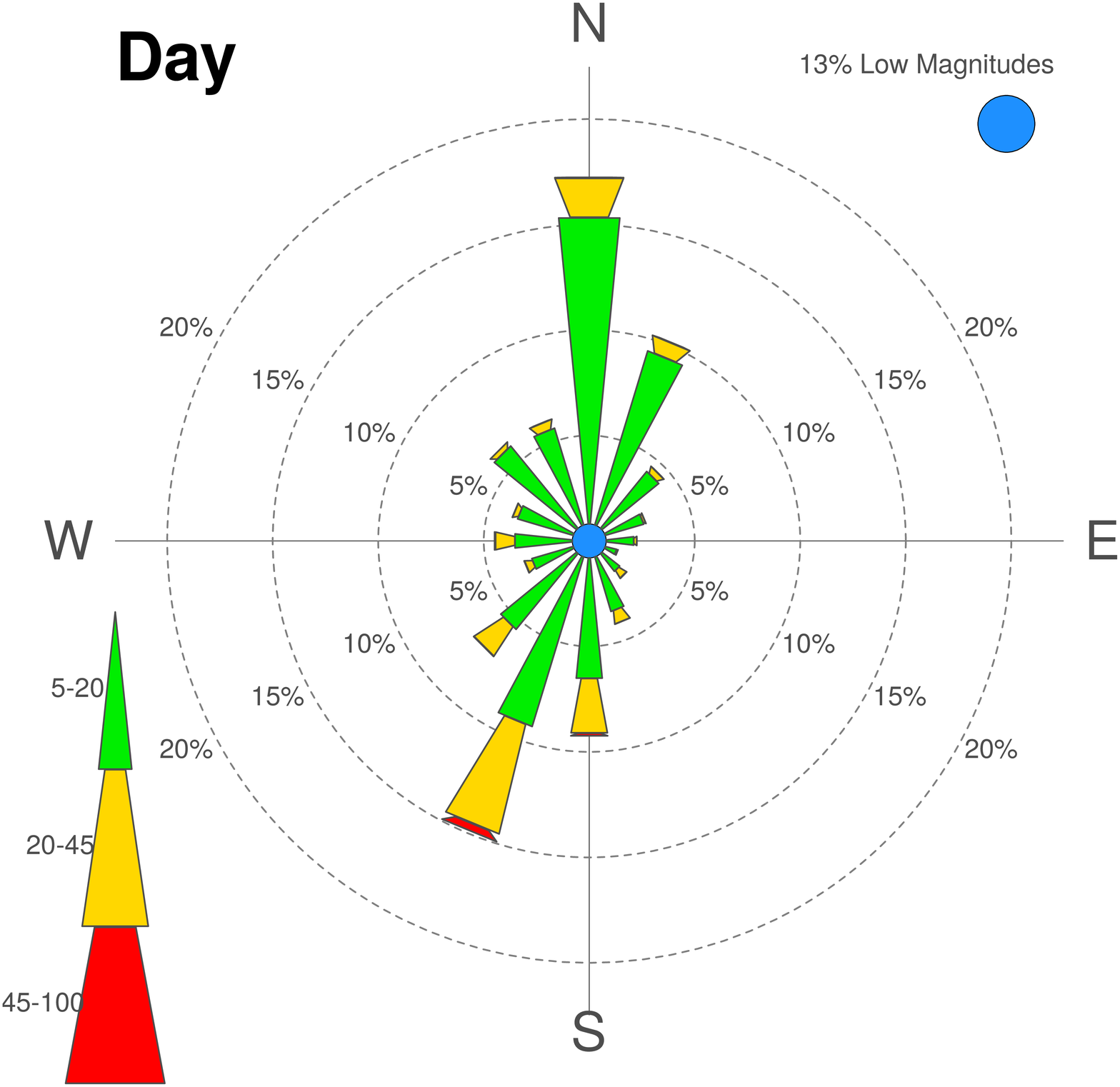}%
     \hfill%
     \includegraphics[width=6cm,height=6cm]{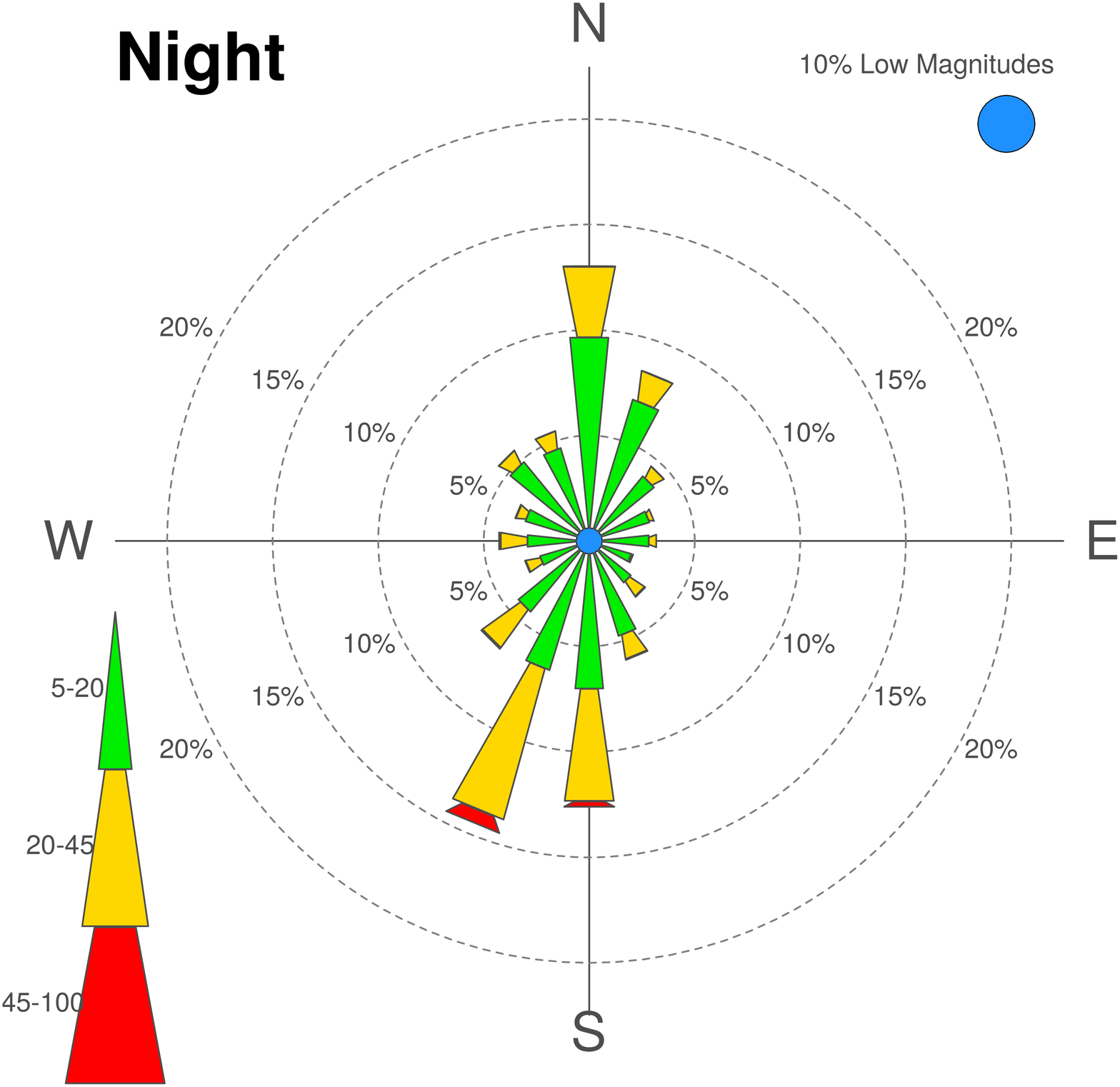}
     }
  \caption{We present wind roses (\textit{a}) for daytime and (\textit{b}) nighttime. 
  The different colors within the roses indicate different velocity ranges.  In the center of the wind roses, 
  ``low magnitudes''   represent velocities in the range (1\,km hr$^{-1}$ -- 5\,km hr$^{-1}$). The highest 
  velocities, 45--100\,km hr$^{-1}$ are represented in red color. The data used here is from sites 1 and 3.}
  \label{fig:winddir2}
\end{figure}

In Figure~\ref{fig:windDiff} we present hexagonal bin plots to show the distribution of values of the 
difference between wind gust and sustained wind speeds as a function of sustained wind speed for daytime and 
nighttime data. The color bar in the right part of each gragh indicates the number of counts in each 
hexagonal bin. From Figure~\ref{fig:windDiff}, it can be seen that given the limit of 45\,km hr$^{-1}$ 
(12.5\,m s$^{-1}$) in sustained wind speed, the corresponding wind gust speed will most probably 
be inside the range of 55--85\,km hr$^{-1}$ (15.3--23.6\,m s$^{-1}$).

Table~\ref{tab:wind} and Figure~\ref{fig:allparam} present the monthly means of the sustained wind 
speed, where it can be seen that the strongest winds occur in  December, January, February and March. 
Table~\ref{tab:highWind} presents the dates with very strong sustained winds, defined as speeds 
$\ge 65$\,km hr$^{-1}$ ($\ge 18$\,m s$^{-1}$) for more than 15 min. Most of the time, these very strong 
winds occur during November to February and come from SSW direction. Surprisingly, most of these events 
occurred during daytime. For these events, wind gust speeds usually are $20-30$\,km hr$^{-1}$ stronger 
than the sustained wind speed. The strongest wind gust speed registered is $129$\,km hr$^{-1}$ 
and it has been recorded four times since 2015, coming from SSW direction at night. 

\begin{table}[!t]\centering
  \setlength{\tabnotewidth}{0.5\columnwidth}
  \tablecols{5}
  \setlength{\tabcolsep}{2\tabcolsep}
  \caption{Monthly mean sustained wind and wind gust speeds} \label{tab:wind}
 \begin{tabular}{lrrrr}
    \toprule
               & \multicolumn{2}{c}{Wind} &  \multicolumn{2}{c}{Gusts} \\
   \cmidrule{2-5}
               & Day & Night  & Day & Night \\
   \cmidrule{2-5}
    Month      &  \multicolumn{4}{c}{(km hr$^{-1}$)} \\
   \cmidrule{1-5}
    January    & $15 \pm 10 $ & $ 17\pm9 $    & $26 \pm 14 $ & $27 \pm14 $  \\
    February   & $17 \pm 10 $ & $ 19\pm11 $   & $29 \pm 15 $ & $30 \pm15 $  \\
    March      & $14 \pm 10 $ & $ 18\pm10 $   & $26 \pm 14 $ & $28 \pm14 $  \\
    April      & $14 \pm 8 $  & $ 17\pm10 $   & $25 \pm 12 $ & $27 \pm14 $  \\
    May        & $13 \pm 8 $  & $ 16\pm10 $   & $23 \pm 11 $ & $25 \pm13 $  \\
    June       & $12 \pm 7 $  & $ 15\pm9 $    & $21 \pm 10 $ & $23 \pm13 $  \\
    July       & $9 \pm 5 $   & $ 9\pm6 $     & $17 \pm 7 $  & $16 \pm8 $  \\
    August     & $9 \pm 5 $   & $ 10\pm6 $    & $17 \pm 7 $  & $17 \pm9 $  \\
    September  & $10 \pm 6 $  & $ 12\pm8 $    & $19 \pm 10 $ & $20 \pm12 $  \\
    October    & $12 \pm 8 $  & $ 15\pm9 $    & $22 \pm 12 $ & $24 \pm12 $  \\
    November   & $13 \pm 9 $  & $ 16\pm10 $   & $23 \pm 13 $ & $25 \pm14 $  \\
    December   & $16 \pm 10 $ & $ 17\pm10 $   & $28 \pm 15 $ & $28 \pm15 $  \\

    \bottomrule
  \end{tabular}
\end{table}

\begin{table}[!ht]\centering
  \setlength{\tabnotewidth}{0.5\columnwidth}
  \tablecols{5}
  \setlength{\tabcolsep}{1\tabcolsep}
  \caption{High wind speeds\tabnotemark{a}} \label{tab:highWind}
  \begin{tabular}{lrrcl}
    \toprule
                  & \multicolumn{1}{c}{Wind}    & Gust  &  Wind        &    Time lapse\tabnotemark{b}\\
                  & \multicolumn{2}{c}{speed}            &   direction  &  \\
   \cmidrule{2-5}
  \multicolumn{1}{c}{Date}     & \multicolumn{2}{c}{(km hr$^{-1}$)}  & &  \\
   \cmidrule{1-5}
   Dec 22, 2009   & $66  $  & $83 $ &    W    & at 9:25--9:50\,hrs \\
   Nov  4, 2011   & $68  $  & $102$ &    SSW  & at 16:25--16:55\,hrs \\
   Jan 23, 2012   & $68  $  & $94 $ &    SSW  & at 14:30--15:05\,hrs \\
   Dec 16, 2016   & $70  $  & $101$ &    S    & at 6:10--7:30\,hrs \\
   Feb  2, 2019   & $67  $  & $96 $ &    SSW  & at 16:00--17:25\,hrs \\
   Feb 14, 2019   & $73  $  & $109$ &    SSW  & at 12:10--17:25\,hrs \\
    \bottomrule

    \tabnotetext{a}{\small Sustained wind speed $\ge 65$\,km hr$^{-1}$  for more than 15 minutes.}
   \end{tabular}
\end{table}

Wind direction statistics are evaluated by calculating the percentage of the time in which the wind comes 
from each direction.  The wind roses have been divided into sixteen mean directions. We have restricted 
sustained wind speed to values larger than or equal to 1\,km hr$^{-1}$ (0.3\,m s$^{-1}$), in order to obtain a set 
of wind direction values that is reliable.  We tabulate lower wind speeds separately, without regard for their direction.

The wind roses of the sustained wind direction for daytime and nighttime are shown in Figures~\ref{fig:winddir2}a 
and \ref{fig:winddir2}b, respectively. From this figure, there is little difference between daytime and 
nighttime in terms of wind direction. Taken as a whole, the predominant wind directions are SSW 
(202$^{\circ}\pm$11$^{\circ}$) and N (0$^{\circ}\pm$11$^{\circ}$). The wind rarely comes from the West or East. 
The strongest winds, those $\ge$45\,km hr$^{-1}$,  only come from  South and SSW directions, while low speed 
winds mostly come from the North. Also, from Figures~\ref{fig:winddir2}a and b it can be seen that high 
speed winds ($\ge$20\,km hr$^{-1}$) are more common at night, while low speed winds ($5-20$\,km hr$^{-1}$) 
come during daytime.

\begin{figure}[!ht]\centering
  \includegraphics[width=0.8\columnwidth]{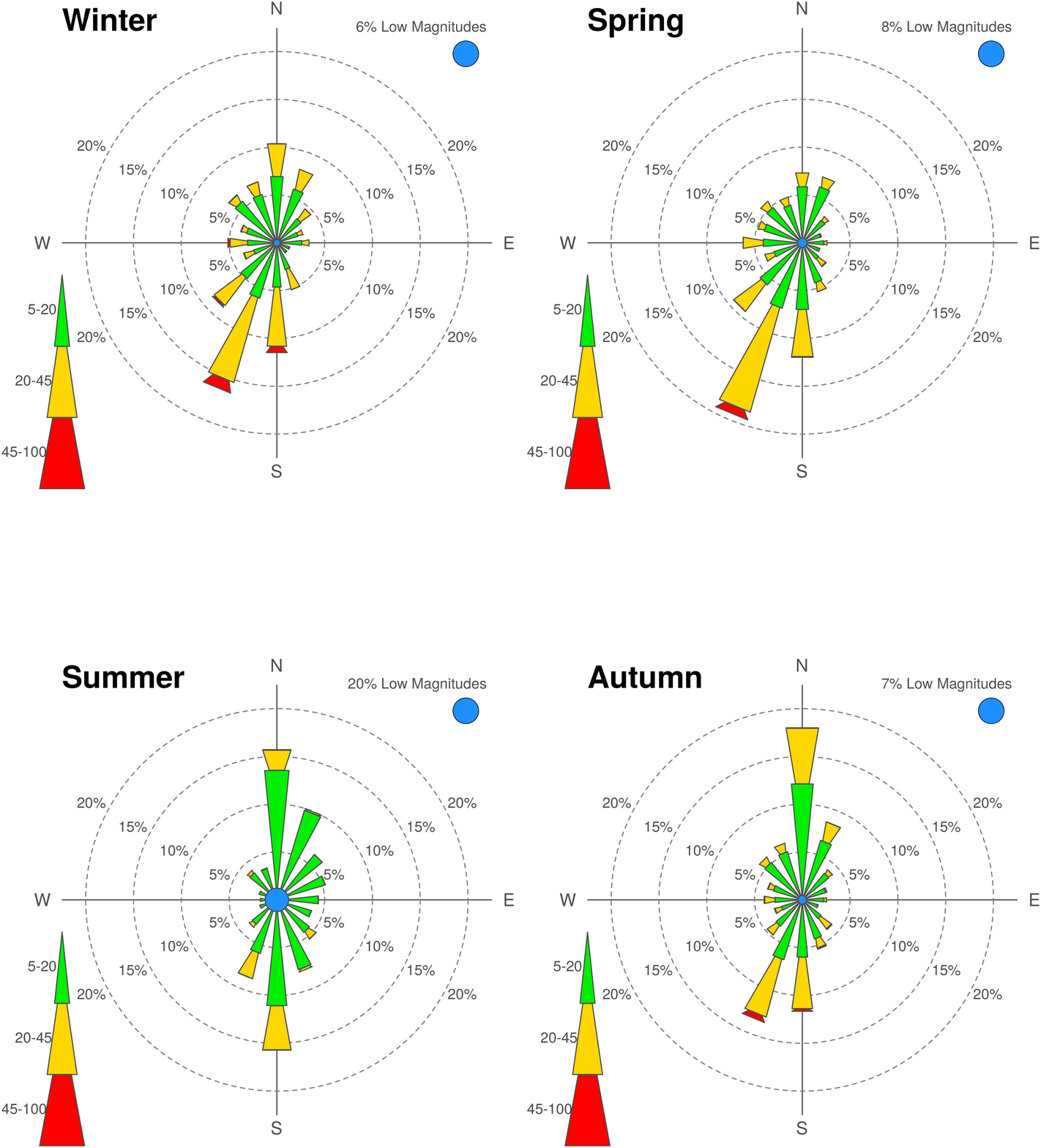}
  \caption{We present nighttime wind roses for each season. 
  The different colors within the roses indicate different velocity ranges. In the center of the wind roses, 
  ``low magnitudes''   represent velocities in the range (1\,km hr$^{-1}$ -- 5\,km hr$^{-1}$). The highest 
  velocities, 45--100\,km hr$^{-1}$ are represented in red color. The data used here is from sites 1 and 3. }
  \label{fig:windDirseason}
\end{figure}

\begin{figure}[!ht]\centering
  \includegraphics[width=1.0\columnwidth]{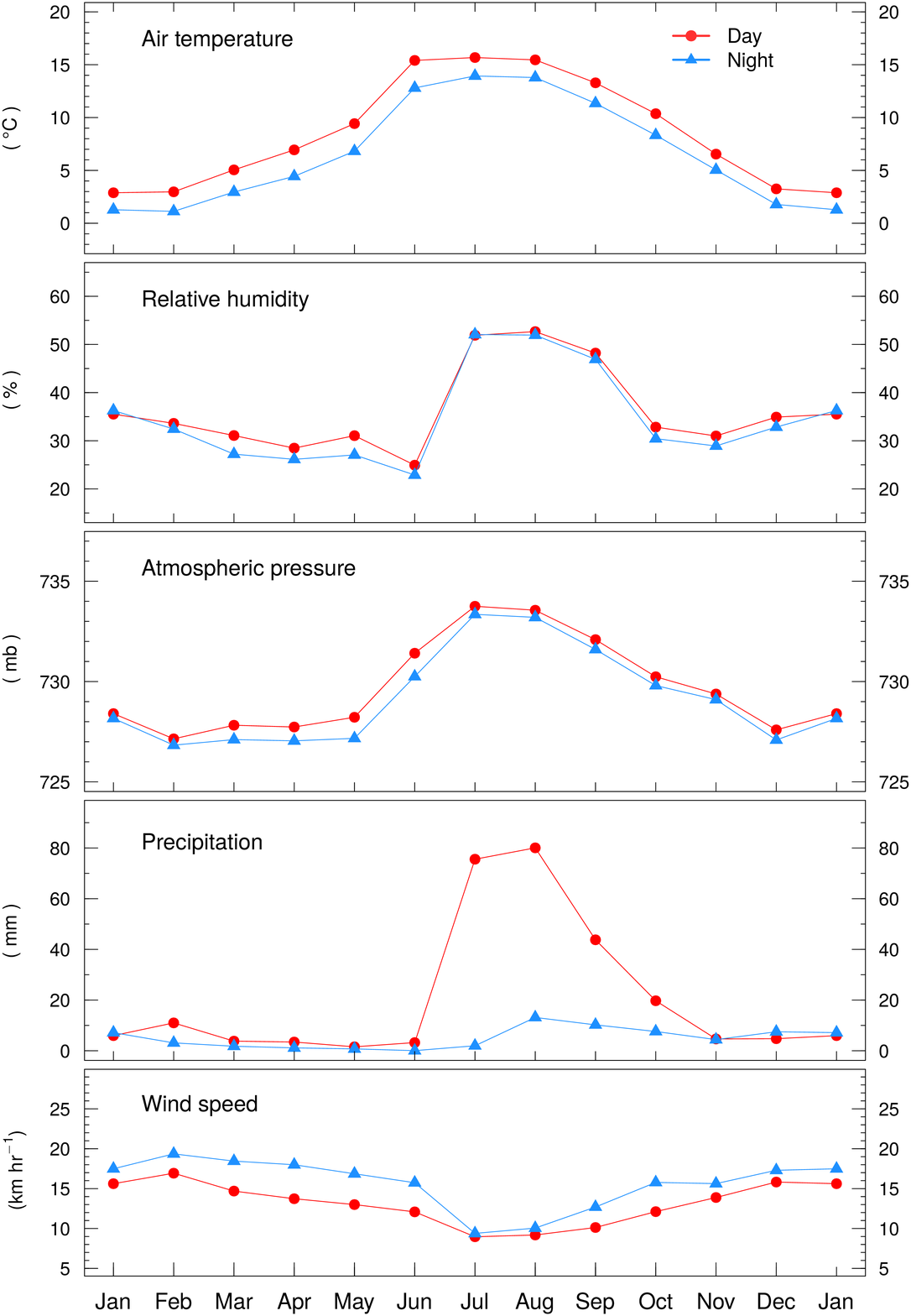}
  \caption{We present the seasonal trends for (top to bottom) the mean air temperature, relative humidity, 
  atmospheric pressure, precipitation, and sustained wind speed for daytime (circles/red) and nighttime (triangles/blue) 
  for the period 2007-2019.}
  \label{fig:allparam}
\end{figure}

Since wind direction is similar for daytime and night data, in the following we present only the nighttime 
wind roses. In Figure~\ref{fig:windDirseason}, we present the wind roses for each season. From this figure, 
it can be seen that during winter and spring the dominant wind direction is SSW direction (202$^{\circ}$) and that 
the strongest winds ($> 45$\,km hr$^{-1}$) also come from this direction. During the summer, there is a 
bipolar distribution of the winds, with North and South dominant directions, but no strong winds. 
Finally, in the autumn, most of the winds come from the North and some strong winds from the South and SSW direcion.

\clearpage
\section{Conclusions}
\label{sec:conclu}

We analyze thirteen years of weather conditions at the Observatorio Astronómico Nacional in the Sierra San 
Pedro Mártir.  The weather parameters we consider are: air temperature, relative humidity, atmospheric pressure, 
precipitation, sustained and wind gust speeds and direction. The data have been split in two sets, one for daytime
and the other for nighttime. Figure~\ref{fig:allparam} summarizes many of our results. Our main conclusions are:

\begin{itemize}
 \item Most of the time, the air temperature is below 15$^{\circ}$C. The median temperature is 10.3$^{\circ}$\,C and 
 7.0$^{\circ}$\,C for daytime and nighttime, respectively, with a median diurnal variation of $\Delta T=5.4^{\circ}$C. 
 At night, the temperature is quite constant with a mean value of the rate of change of $-0.06\pm 0.22^{\circ}$C 
 per hour around midnight. The coldest month is January with a mean temperature of 2.1$^{\circ}$\,C and the warmest 
 month is July with a mean temperature of 15.1$^{\circ}$\,C. We searched for an increase in temperature over the last 
 thirteen years and over the last  50 years (using data from \citealp{2007RMxAC..31..113A}), but have not found any 
 significant variation, at least at the 95\% confidence level. Also, we have not found any increase in the number 
 of warm days nor a decrease in the number of cold days over the thirteen years spanned by our data.
 
 \item The relative humidity is statistically different for daytime and nighttime with median values of 30\,\% 
 and 27\,\%, respectively.  The diurnal variation of this parameter shows that the relative humidity is quite 
 constant from midnight to sunrise, but that it varies during the rest of the day.  The largest variations are 
 found in spring and summer. The driest season is spring with a median relative humidity of 23\,\% while the most 
 humid season is summer with median relative humidity of 51\,\%. We find no evidence for long term evolution 
 in the relative humidity over the time span of our data.

  \item The median value of the atmospheric pressure is $\sim$730\,mb.  During daytime, the atmospheric pressure 
  is slightly higher than at night, by 0.5\,mb. This parameter varies seasonally, with the highest atmospheric 
  pressure during summer. The diurnal variation of the atmospheric pressure is $\sim2$\,mb, due to the phenomenon 
  known as atmospheric tides. On average, the atmospheric pressure has increased by 1\,mb in the last decade 
  (95\,\% confidence level).
  
 \item The annual mean accumulated precipitation is $\sim$313\,mm, 70\,\% of which occurs during summer time. 
 Also, on average it rains four times more during the day than at night. However, our precipitation data 
 for winter is an under-estimate of the true precipitation due to insensitivity of our sensors 
 to frozen  precipitation. Correcting for this, we estimate that the total annual precipitation at the OAN-SPM is 
 400-450\, mm. The annual accumulated precipitation has increased over the time span of our data, with 30\,mm 
 more precipitation every year since 2009 (99\,\% confidence level).
 
 \item The median value of the sustained wind speed is 11\,km hr$^{-1}$ during the day and 13\,km hr$^{-1}$ at 
 night. The predominant wind directions are SSW (202$^{\circ}$) and North (0$^{\circ}$). The sustained wind is 
 stronger at night, during the winter, and comes mainly from SSW direction. The median value of the wind gust 
 speed are 20\,km hr$^{-1}$ and 22\,km hr$^{-1}$ for daytime and nighttime, respectively. Wind gust speeds are 
 9\,km hr$^{-1}$ stronger than sustained wind speeds 50\% of the time. Only 5\% of the time the diferrence between 
 wind gust and sustained wind speeds exceeds 20\,km hr$^{-1}$.

 \end{itemize}

\acknowledgments
{\it Acknowledgments:}
We thank the anonymous referee for the cafeful reading and time spent to analyze this manuscript and the 
constructive comments and suggestions. This work is based upon observations carried out at the Observatorio 
Astronómico Nacional on the Sierra San Pedro Mártir (OAN-SPM), Baja California, México. We thank the daytime 
and night support staff at the OAN-SPM for facilitating and helping obtain these data: E. Cadena, T. Calvario, 
U. Ceseña, A. Córdova, A. Franco, B. García, F. Guillén, G. Guisa, J. Hernández,  J. Herrera, D. Hiriart, 
E. López, B. Martínez, G. Melgoza, F. Montalvo, S. Monrroy, F. Murillo, C. Narváez, M. Núñez, J.~L. Ochoa, 
F. Quiros, M. Reyes, H. Serrano, E. Valdés, J. Valdez, F. Valenzuela, I. Zavala and S. Zazueta. This paper is 
dedicated to the memory of our friends Gabriel Leyva Arias y Roberto Higuera Escalante.
  
Finally I. P-F. wants to acknowledge the use of R package, which is a free software environment for 
statistical computing and graphics\footnote{\texttt{http://www.r-project.org}}.

\end{document}